\documentstyle[12pt,aas2pp4]{article}
\begin{document}
\title{A Spectroscopic Survey of the Galaxy Cluster \\ CL 1358+62 at
$z$=0.328}
\author{David Fisher\altaffilmark{1}}
\affil{Kapteyn Astronomical Institute, University of Groningen, Postbus 800,
9700 AV Groningen, The Netherlands}
\authoremail{fish@astro.rug.nl}
\author{Daniel Fabricant}
\affil{Harvard-Smithsonian Center for Astrophysics, 60 Garden Street,
Cambridge, MA 02138}
\authoremail{dgf@spectra.harvard.edu}
\author{Marijn Franx and Pieter van Dokkum}
\affil{Kapteyn Astronomical Institute, University of Groningen, Postbus 800, 
9700 AV Groningen, The Netherlands}
\authoremail{franx@astro.rug.nl and dokkum@astro.rug.nl}
\vskip 0.1in
\centerline{\it To appear in the 1998 May 1 Astrophysical Journal}
\altaffiltext{1}{Current address: Universit$\ddot{\rm a}$t Hamburg, 
Hamburger Sternwarte, Gojenbergsweg 112, 21029 Hamburg, Germany}
\begin{abstract}
We present a spectroscopic survey of the rich, X-ray selected, galaxy cluster
CL 1358+6245 at $z$=0.328. When our 173 new multi-slit spectra of cluster
galaxies are combined with data from the literature, we produce a catalog of
232 cluster members in a region 10\arcmin~$\times$ 11\arcmin~(3.5 Mpc $\times$
3.8 Mpc) surrounding the brightest cluster galaxy. These data are used to study
the structure and dynamics of the cluster and to examine the radial and
velocity distributions as a function of spectral type. We classify the spectral
types of the cluster members according to the strengths of the Balmer
absorption lines (H$\delta$, H$\gamma$, and H$\beta$) and the [OII]
3727\thinspace\AA~emission line.

We derive a mean redshift of $z$=0.3283$\pm$0.0003, and a velocity dispersion
of 1027$^{+51}_{-45}$ km s$^{-1}$ for the 232 cluster members. However, the
cluster velocity distribution is non-Gaussian and we identify at least two
subgroups with 10-20 members and dispersions $<$400 km s$^{-1}$. The fraction
of spectroscopically active galaxies (poststarburst and emission-line) in the
core of the velocity distribution (within 0.6 $\sigma$ of the mean cluster
velocity) is 16$\pm$4\%, rising to 32$\pm$7\% for galaxies in the tails of the
velocity distribution (2.0 $\sigma$ from the mean cluster velocity). In total,
the cluster is composed of 19$\pm$3\% emission-line and 5$\pm$1\% poststarburst
galaxies. The velocity dispersion of the poststarburst galaxies is close to
$2^{1/2}$ times that of the absorption-line galaxies, consistent with free-fall
accretion of the poststarburst galaxies.

The changing mix of galaxy spectral types as a function of local galaxy density
(or distance from the cluster center) in CL 1358+62 is similar to what is
observed in nearby rich clusters. The percentage of emission-line galaxies
increases steadily with radius from 9$\pm$3\% within a radius of
$r_{\rm p}$$<$0.7 Mpc, to 41$\pm$9\% in our outermost radial bin at $\sim$1.7
Mpc. The percentage of absorption-line galaxies falls from 84$\pm$9\% to
59$\pm$11\% over the same radial intervals. These results are consistent with
the idea that the cluster grows through the accretion of late-type field
galaxies, a fraction of which are transformed into poststarburst galaxies
during the accretion event.

Our high signal-to-noise spectra of the ``E+A'' galaxies allow a detailed
comparison with spectra of nearby merging and strongly interacting galaxy
systems. We find that nearby mergers have stronger [OII]
3727\thinspace\AA~emission ([OII] EW $>$ 5\thinspace\AA) than that observed for
our ``E+A'' galaxies. This implies either that the ``E+A''galaxies in CL
1358+62 were not formed through major mergers, or that if they were formed via
merging, their gas-supply was then quickly depleted, possibly by the ICM.

\end{abstract}


\section{Introduction}
CL 1358+6245 is one of the richest medium redshift galaxy clusters
known (see Luppino et al.~1991). The cluster was originally selected
from the {\it Einstein} Medium Sensitivity Survey (Gioia et al.~1990,
Gioia \& Luppino 1994) by Fabricant, McClintock \& Bautz (1991,
hereafter FMB) on the basis of its large X-ray luminosity:
$L_x$$\sim$7$\times$ 10$^{44}$ ergs s$^{-1}$ (0.2-4.5 keV). The FMB
study used X-ray selection to avoid possible biases associated with
optical selection (see Koo 1988), although they established that the
galaxy populations of X-ray selected clusters at intermediate
redshifts are similar to those in optically selected clusters.

Here, we present a much expanded spectroscopic study of the cluster
based on new data obtained at the William Herschel Telescope (WHT)
with the Low Dispersion Survey Spectrograph (LDSS-2). We obtained
redshifts of 328 additional galaxies, 173 of which are cluster
members. The spectra are of sufficient quality to determine the main
characteristics of the stellar populations in the cluster galaxies
from line strength measurements of emission and absorption
features. We use the spectroscopic data to explore the cluster dynamics
and the evolution of its galaxy population. The work reported here is
part of a larger study of the evolution of X-ray selected clusters at
increasing redshift, combining ground-based spectroscopy and
photometry with HST WFPC2 images.

There is now substantial evidence that galaxy populations in distant
clusters differ from the populations in nearby clusters. Butcher \&
Oemler (1978, 1984) demonstrated that the fraction of blue galaxies in
clusters increases from about 3\% in nearby clusters to 25\% at
$z$$\sim$0.5. Spectra of these blue galaxies revealed both
emission-line galaxies, and surprisingly, poststarburst galaxies with
strong Balmer absorption lines (Dressler \& Gunn 1983). The
poststarburst galaxies (often called ``E+A'' galaxies) are believed
to have experienced a burst of star formation that ended approximately
1 Gyr before the epoch of observation.

Previous studies have suggested that the emission-line and
poststarburst galaxies are found preferentially in the outer regions
of clusters (e.g., Couch \& Sharples 1987; FMB; Colless \& Dunn 1996;
Abraham et al.~1996). A goal of our current program is to map the
distribution of the spectroscopically active galaxies to help
constrain their evolutionary history. HST studies will reveal the
morphology of these distant galaxies, and will add important
evolutionary constraints, but ground-based spectroscopy of the type we
report here remains essential to our progress in this field.

In addition to studying the evolution of cluster galaxies, we wish to
address the evolution of clusters as dynamical systems. Simulations
(reviewed in Evrard 1993) suggest that in bottom-up hierarchical
clustering models, many generations of subcluster merging are likely
to have occurred in rich clusters by the present epoch. Furthermore,
these models predict that the accretion of galaxies into the cluster
and the frequency of localized velocity structures should increase as
a function of lookback time (Gunn \& Gott 1972). Large numbers of
redshifts are needed to determine the degree of dynamical mixing
within a cluster and observations at the current epoch have yielded
important tests of these models of cluster formation. Observations at
intermediate redshift have only recently become available (e.g.,
Carlberg 1994) and are of particular importance because they can be
used to directly search for dynamical evolution.

The paper is organized as follows. First, we describe the photometric
catalog (\S 2) from which we selected objects for spectroscopy. The
design of the multislit aperture masks used to obtain the
spectroscopy, spectroscopic reduction techniques, and determination of
radial velocities are discussed in \S 3. The spectroscopic catalog is
presented in \S 4 including a description of the completeness of the
spectroscopic sample and the definition of the line strength indices
used to classify the galaxy spectra. In \S 5, we discuss the spectral
types observed in our galaxy sample. The changing mix of spectral
types as a function of local galaxy density or distance from the
cluster center and the dependence of the galaxy velocity distribution
on spectral type are discussed in \S 6. Our search for subclustering
is discussed in \S 7. A summary of our findings is presented in
Section \S 8.

\section{Photometric Catalog}
In 1992 we obtained $V$ and $R$ images of CL 1358+62 at the Fred
L.~Whipple Observatory's 1.2-m telescope, covering the central
10\arcmin~$\times$ 11\arcmin~portion of the cluster. Multiple
exposures in $V$ and $R$ were obtained with small spatial offsets
between exposures. The observations consist of three 900 sec $R$
exposures and five 1200 sec $V$ exposures. We used a Ford
2048$\times$2048 pixel CCD with a scale of 0.325\arcsec~pixel$^{-1}$.
The seeing was 1.2\arcsec~during the run
and conditions were not photometric so we normalized to the FMB
zeropoints for $V$ and $R$. The images were reduced in the standard
fashion to create a master frame for each run and each filter. We used
the Faint Object Classification and Analysis System (Jarvis \& Tyson
1981; Valdes 1982a,b, hereafter FOCAS) to extract aperture photometry,
positions and object classifications (galaxy, star, defect). We used
10 pixel diameter (3.25\arcsec) apertures for the data, close to the
3.3\arcsec~apertures that FMB used. At the distance of CL 1358+62,
3.3\arcsec~corresponds to 19 kpc (for consistency with FMB we assume
H$_0$=50 km s$^{-1}$ Mpc$^{-1}$ and q$_0$=1/2 throughout). The
expected reddening towards CL 1358+62 is small [E($B-V$)=0.01] and so
no corrections have been made for Galactic reddening (Burstein \&
Heiles 1984). The photometric catalog resulting from these images was
used to select objects for our spectroscopic runs.

In 1994, we obtained additional $V$ and $R$ images with the WHT LDSS-2
spectrograph (operated as a focal reducer) that cover an approximately
circular region of 8\arcmin~diameter slightly offset from the cluster
center. The LDSS-2 detector was a 1024$\times$1024 Tektronix CCD with
a scale of 0.592\arcsec~pixel$^{-1}$. The WHT observations consist of
seven 300 sec $R$ exposures and two 300 sec $V$ exposures. The seeing
was 1.6\arcsec~during the run and flux standards were not observed so
we extracted counts in $\sim$3.3\arcsec~apertures and normalized to
the FMB zeropoints. The LDSS-2 images are rather coarsely sampled, but
are deeper than the Whipple 1.2-m images.  Aperture photometry in 6
pixel diameter (3.55\arcsec) apertures was extracted from the reduced
images with the FOCAS package.  The $V-R$ object colors presented here
are drawn from this WHT catalog.  

We can estimate the completeness of the WHT photometry by examining
the distribution of galaxy counts as a function of magnitude well away
from the cluster core.  We find that log N$\propto$$R^{0.3}$ for
20$<R<$23.5.  The counts in the 23.5$<R<$24 bin fall 20\% below this
relation and the counts in the 24$<R<$24.5 are over a factor of two
below the relation.  We therefore expect that the WHT photometry is
substantially complete to $R\sim$23.5.  There are two areas of local
incompleteness surrounding two stars that saturated the CCD: one of
radius $\sim$50$^{\prime\prime}$ at 13$^h$ 58$^m$ 21.0$^s$,
62$^{\circ}$ 50$^{\prime}$ 23$^{\prime\prime}$ (1950) and one of
radius $\sim$20$^{\prime\prime}$ at 13$^h$ 58$^m$ 52.1$^s$,
62$^{\circ}$ 44$^{\prime}$ 56$^{\prime\prime}$ (1950).

A multi-color mosaic of HST WFPC2 images of CL 1358 has been obtained.
This mosaic offers an unprecedented high resolution view of a large
area of a cluster at intermediate redshift.  We refer the reader to a
companion paper (van Dokkum et al.~1997), which uses these HST images
to study the galaxy morphologies and color-magnitude relation in the
cluster.

\section{WHT Spectroscopy}
The new spectra were obtained during two observing runs with the
LDSS-2 spectrograph at the WHT: a 3 night run in April 1994 and a 5
night run in June 1996. The LDSS-2 is a multi-slit spectrograph with
an 11\arcmin~$\times$11\arcmin~field of view that is mounted at the
Cassegrain focus of the WHT (see Allington-Smith et al.~1994). The
aperture plates used for our two runs contained slitlets
1.5\arcsec~wide and $\sim$15\arcsec~long.  The masks used for the
first WHT run each contained $\sim$30 slitlets; the masks for the
second run each contained $\sim$50 slitlets.

\subsection{Aperture Plate Layout}
The design of aperture plates for the MMT is described in FMB.  FMB
obtained spectra for 70 galaxies within $R_{30}$ ($\sim$2\arcmin) of
the dominant central cluster galaxy, yielding 59 redshifts of cluster
members.

For the first WHT spectroscopic run galaxies with $R \le$ 21 were
given highest priority.  Nine masks were designed with an optimizer
which attempted to maximize the assignment of high priority galaxies.
Fainter galaxies were assigned priorities proportional to their
magnitude and were used to fill the remaining spaces in the
masks. Observing time was available to expose six of the nine masks,
yielding spectra of 127 high priority galaxies and 35 lower priority
galaxies.

Similar galaxy priorities were assigned for the second WHT run except
that objects falling outside the HST mosaic (8\arcmin $\times$
8\arcmin) were assigned the lowest priority independent of their
magnitude. The automated selection procedure was used to design nine
masks and all nine aperture plates were exposed.  Spectra for 42
galaxies with $R \le$ 21 and 172 fainter galaxies were obtained.

\subsection{Instrument Configuration}
For the 1994 WHT run and the first 2 nights of the 1996 WHT run, we
used a Tektronix 1024$\times$1024 CCD with 24 $\mu$m pixels with a
pixel scale of 0.59\arcsec~pixel$^{-1}$. For the last 3 nights of the
1996 WHT run we used a Loral 2048$\times$2048 CCD with 15 $\mu$m
pixels and a spatial scale of 0.37\arcsec~pixel$^{-1}$. The read out
noise and gain of the Tektronic chip was 4.4 e$^-$ and 1.2 e$^-$/ADU
respectively, while that of the Loral chip was 6 e$^-$ and 0.8
e$^-$/ADU. Exposure times were 4$\times$1800 sec for the Tektronic
chip and 3$\times$2400 sec for the Loral with each aperture plate.

The 300 line mm$^{-1}$ grism blazed at 5000 \AA~was used for all
observations, providing a resolution of 13 \AA~FWHM with the aperture
plates.  The dispersion was 5.3 \AA~pixel$^{-1}$ with the Tektronix CCD
and 3.3 \AA~pixel$^{-1}$ with the Loral CCD.  The wavelength coverage
depends on the layout of the aperture mask but typically includes the
range 3500-8500 \AA.  This wide wavelength coverage is beneficial for
the identification of field galaxies to $z \sim$1.  In the cluster rest
frame our coverage extends from well below the [OII]
3727\thinspace\AA~emission line to well beyond the Mg I triplet at
5175\thinspace\AA. 

\subsection{Data Reduction and Analysis}
Data reduction for the 1994 WHT run was carried out with IRAF while
reduction of the 1996 WHT run was carried out with VISTA (Lauer,
Stover, \& Terndrup 1983).  In brief, the processing steps for each
CCD exposure were: bias subtraction, flat-fielding, wavelength
calibration and geometrical rectification. Following this, cosmic ray
events were removed and the individual spectra were extracted.
One-dimensional spectra, binned logarithmically in the wavelength
direction, were the final product of these procedures. Flux standards
were not obtained, so flux calibration was not attempted.

The identification and removal of cosmic ray events is a critical step
in the reduction process because they can be confused with emission
lines.  We generated a cosmic ray map for each frame by subtracting a
median of the multiple exposures of each mask from the individual
exposures and setting to zero all pixels less than 5$\sigma$ above the
median level.  Following subtraction of the cosmic ray map, each frame
was visually inspected for remaining deviant pixels before combining
it with the other exposures.  Straight medians of the individual
exposures were created to aid the separation of cosmic ray events from
emission lines in ambiguous cases.

The major difference between the reduction procedures for the two runs
was that the wavelength calibration was performed after extraction of
one dimensional spectra for the 1994 data, while it was performed
prior to extraction for the 1996 data. The results of both procedures
are essentially indistinguishable. CuAr spectra were obtained with
each aperture plate to determine the wavelength solutions for each
slitlet.  Wavelength calibration for each slitlet was performed using
a fifth-order polynomial fit to $\sim$15-20 arc spectrum lines. The
rms residuals from these fits are $\lesssim$0.2 \AA. As a final step
in the wavelength calibration the position of the 5577 \AA~night
skyline was measured and used to correct for any offset to the
wavelength solution caused by relative shifts between the arc and
galaxy exposures. This correction was usually of order 1 \AA~or less.

We estimate that residual errors in the final wavelength-calibrated
and line curvature-corrected spectra are present at the level of
$\lesssim$0.5\thinspace\AA, as determined from measurements of the
positions of prominent spectral features. By themselves, these errors
can result in redshift uncertainties of $\sim$28 km s$^{-1}$ or
$\varepsilon$($z$)$\approx$0.00009.  We discuss our estimated external
errors below.

\subsection{Determination of Redshifts}
Redshifts were derived from either the absorption lines using a
cross-correlation technique (Tonry \& Davis 1979) or directly from the
emission lines. These techniques are similar to those used by FMB. The
subsections below describe the details of the derivations of radial
velocities for the cluster members.

\subsubsection{Absorption Line Redshifts}
Redshifts were determined by cross-correlating template spectra with
cluster galaxy spectra for the 188 (177 red absorption-line galaxies
$+$ 11 poststarburst galaxies) galaxies with WHT spectra dominated by
stellar absorption features.  Two high signal-to-noise template
spectra were used: a spectrum of the central regions of the early-type
galaxy NGC 7331 and a spectrum of a poststarburst galaxy in the
cluster (number 209). A semiautomated procedure was developed to
cross-correlate the object spectra with the templates using the
cross-correlation routine, XCSAO.  XCSAO (Kurtz et al.~1992) is part
of the RVSAO package for the IRAF environment. The velocities and
their errors are returned by the XCSAO software. Each object spectrum
was cross-correlated with the each of the two templates within the
3530$-$5110\thinspace\AA~rest frame wavelength range.

The 14 galaxies from the FMB study which were reobserved with the WHT
allow us to estimate the external errors.  The RMS velocity difference
between the FMB and the new WHT results is $\sim$100 km s$^{-1}$,
yielding an estimated uncertainty of $\sim$70 km s$^{-1}$ for an
individual measurement if the MMT and WHT errors are equal.  In fact,
the WHT have higher resolution and (in general) higher signal-to-noise. 
We therefore assign an error of 60 km s$^{-1}$ to the highest quality
WHT spectra and 80 km s$^{-1}$ to the highest quality MMT spectra.  We
estimate that the errors in the lower quality MMT spectra are 120 km
s$^{-1}$. This comparison of the FMB and WHT redshifts suggests
that FMB significantly overestimated their errors. 

\subsubsection{Emission Line Redshifts} 
The sky subtracted
two-dimensional slitlet frames and one-dimensional object spectra were
inspected for emission features and they were detected in 44 galaxies. 
For these galaxies redshifts were often determined solely from the [OII]
3727\thinspace\AA, although in some cases additional lines (usually
[OIII] 5007\thinspace\AA~and H$\beta$) were present.  The emission line
positions and their associated errors were measured by fitting a
Gaussian profile with the IRAF SPLOT task. 

Eight of the galaxy spectra in our catalog have redshifts determined
from both absorption and emission lines.  The largest difference between
the emission-line and absorption-line derived radial velocities for the
eight galaxies is 143 km s$^{-1}$.  The RMS velocity difference for the
eight galaxies is 53 km s$^{-1}$, which implies an internal error in the
WHT velocity determinations of $\sim$40 km s$^{-1}$. 

\section{Spectroscopic Catalog}
Cluster membership was defined to include all galaxies within the
redshift interval 0.31461 $<$ $z$ $<$ 0.34201.  This corresponds to the
4$\sigma_{\rm cl}$ boundary around the cluster mean.  The nearest
galaxies in redshift space outside these boundaries are at
$+$4.5$\sigma_{\rm cl}$ and $+$5.9$\sigma_{\rm cl}$.  Both these
galaxies are also at large projected distance from the cluster center
(260\arcsec, 1.5 Mpc and 310\arcsec, 1.8 Mpc respectively) and are
therefore not likely to be cluster members.  Table 1 presents our
results for the full sample of 232 cluster members including the FMB
results. The galaxy ($x$,$y$) positions are given with respect to the brightest
cluster galaxy (number 375 at R.A.~(1950) 13$^h$58$^m$20$^s$.7 DEC~(1950) 
62$^\circ$45$^{\prime}$33$^{\prime\prime}$) with positive $x$ towards the West
and positive $y$ to the North.

\subsection{Completeness}
The three major selection effects which could influence the
completeness of our survey are uneven spatial sampling, a magnitude
bias, and a color or spectral type bias. A spatial sampling bias could
result from an uneven distribution of the aperture plate slitlets. A
magnitude bias naturally arises because redshifts are more difficult
to measure for faint galaxies.  Completeness may depend on the galaxy
colors or spectral types because redshifts are generally easier to
measure for galaxies with emission lines.  (see, e.g., Yee, Ellingson,
\& Carlberg 1996).

The spatial distribution of all galaxies from the 1.2 m photometric
catalog with $R$$\leq$21 is plotted in Figure 1. The subset of
galaxies for which we have measured redshifts have been marked. The
region covered by the spectroscopy is outlined, i.e.,
$-$5\arcmin$\lesssim$ $x$ $\lesssim$$+$5\arcmin~and
$-$6\arcmin$\lesssim$ $y$ $\lesssim$$+$5\arcmin. Of the 226 galaxies
in this field with R$\leq$21, we have obtained redshifts for 189
(84\%). There are 154 galaxies with $R$$\leq$21 in the inner half of
the spectroscopic field (by area), and we obtained redshifts for 138
(90\%). The outer half of our field (by area) contains 72 galaxies for
which we acquired 51 redshifts (71\%).

The completeness of our spectroscopic sample as a function of
magnitude is shown in Figure 2. Fig.~2 plots the ratio of the number
of galaxies (including cluster and field galaxies) with
measured redshifts to the total number of galaxies in the photometric
catalog in a magnitude bin centered on each galaxy. We used a
magnitude bin of $\pm$0.50 for $R$$\leq$20 and a magnitude bin of
$\pm$0.25 for fainter objects. Those objects brighter than the
brightest cluster galaxy (at $R$=19.09) are foreground
galaxies. Fig.~2 indicates that our spectroscopic sample is $>$80\%
complete for $R$$\leq$21 and drops steeply to 20\% completeness at
R$\sim$22.

To check if our ability to measure redshifts depends on color, we
divided our sample into red ($V-R$$>$0.75) and blue ($V-R$$<$0.75)
samples.  We find that we are 100\% successful in determining
redshifts for objects with $R$$<$21, and at R$\sim$23 the success
ratio drops to 50\%. The success ratio does not depend strongly on
color for R$<$23.5. At fainter magnitudes the statistics are poor: we
detect 1 of 6 targeted blue galaxies and 2 of 4 targeted red galaxies.
We conclude that any color biases (and by inference spectral type
biases) in our catalog are minor.

\subsection{Absorption and Emission Line Strength Measurements}
The resolution and limited signal-to-noise ratio of our spectra
prevent us from studying subtle spectral features.  Therefore, we
concentrate on the most prominent spectral lines that measure star
formation activity: the Balmer absorption lines (H$\delta$, H$\gamma$,
and H$\beta$), and the [OII] 3727\thinspace\AA~emission line.  

Three wavelength regions need to be defined in order to measure a
spectral index: an interval covering the feature of interest, and a
pair of continuum bandpasses on either side of the feature. The
strength of each spectral index is a measure of the flux in the
central bandpass compared to a continuum level set by the
sidebands. The indices we measure here are tabulated as equivalent
widths in $\rm \AA$. These equivalent width (EW) measurements using the
observed spectra are corrected into the rest frame by dividing by
(1+$z$).

The line strength indices that we use are defined in Table 2. The
H$\delta$, H$\gamma$, and H$\beta$ indices are based on Jones \&
Worthey (1995) but we have broadened the continuum sidebands and moved
them further from the central bandpass.  We have moved the Jones \&
Worthey Balmer sidebands because their sidebands overlap the
absorption lines for poststarburst galaxies.  Figure 3 shows the
spectrum of our strongest poststarburst galaxy (number 243) with our
Balmer line index definitions overlaid. The Jones \& Worthy Balmer
indices are located $\sim$20-30 \AA~closer to the feature bandpasses
on both sides, and fall inside the features being measured. We have,
however, used the wavelength regions of Jones \& Worthey for the
central bandpasses because 95\% of the cluster galaxies are {\it not}
poststarburst and wider feature bandpasses for these galaxies would
include unwanted regions of the spectrum in addition to the Balmer
lines. Conversely, for the strongest poststarburst galaxies, the
Jones \& Worthey central bandpasses are too narrow to include the full
widths of the Balmer lines, leading to up to 30\% underestimates of the
true equivalent widths. As an example, the poststarburst galaxy shown in
Fig.~3 has an average balmer line strength of 9.1\thinspace\AA~in our
system whereas a measurement including the full line widths gives
12.0\thinspace\AA. For the purposes of spectral
classifications our system provides a rigorous means of measuring the
Balmer line strengths for the wide range of galaxy spectra found in CL
1358+62.

We also define our own [OII] emission-line index which we give in
Table 2. The equivalent widths that we measure for all galaxies
are given in Table 1.  Errors are calculated by measuring the noise in
the continuum sidebands and scaling this to the indices themselves.

\section{Galaxy Spectral Types}

\subsection{Definition of Spectral Types}
Our spectra display a wide variety of nebular emission line and
absorption line strengths. The galaxy spectra can be sorted into
spectral classes which are directly related to the star formation
histories of the galaxies. We have divided the cluster galaxy spectra
into three spectral types: (1) a population displaying the normal
absorption features of elliptical and S0 galaxies and no emission
lines, (2) poststarburst galaxies with strong Balmer absorption lines
and no emission features, and (3) galaxies with emission lines, a
broad classification which includes blue galaxies as well as red
galaxies with absorption-line spectra and weak emission features.

The dominant spectral type in our catalog corresponds to the first of
these types: spectra of red galaxies that likely have elliptical and
S0 morphologies. Of the 232 cluster members, 177 (76\%) have pure
absorption-line galaxy spectra dominated by the 4000 \AA~break. These
galaxies are labeled ``abs'' in Table 1. The lower right panel of
Figure 4 shows a typical absorption-line spectrum (galaxy number 353).

The spectra of poststarburst galaxies can be modeled as the
combination of an old stellar population whose light is dominated by K
giant stars and a young population of hot, strong Balmer-lined A stars
(Dressler \& Gunn 1983; Couch \& Sharples 1987). We have defined a
sample of 11 poststarburst galaxies which have average Balmer
absorption line strengths, (H$\delta$+H$\gamma$+H$\beta$)/3, greater
than 4.0\thinspace\AA~and [OII] 3727\thinspace\AA~emission less than
5\thinspace\AA. The Balmer line strength measurements are required to be
significant at least at the 1 $\sigma$ level. The 11 poststarburst galaxies
are labeled ``e$+$a'' in Table 1.

Histograms of average Balmer line strength for all 232 cluster members
and for the galaxies without [OII]\thinspace3727\AA~emission are
plotted in Figure 5.  Our cutoff for poststarburst classification
is justified in the lower panel of Figure 6, which is a plot of
average Balmer line strength versus galaxy color. The poststarburst
galaxies can be clearly distinguished from the red absorption-line and
bluer emission-line galaxies. The poststarburst galaxies are blue,
although not as blue as the the galaxies with strong emission lines.

Some of our poststarburst galaxies display very strong Balmer lines
($<$H$\delta$+H$\gamma$+H$\beta$$>$ $\gtrsim$ 8\thinspace\AA),
indicating recent, massive bursts of star formation.  Models have
indicated that star bursts comprising upwards of 20\% of the mass of the
galaxy are required to achieve such large Balmer line strengths
(Dressler \& Gunn 1983, Poggianti \& Barbaro 1996).  The lack of
detectable emission features in these objects indicates that the star
burst was highly efficient in converting gas to stars.  The dominant A
type spectrum also implies that the burst occurred approximately 1 Gyr
before the epoch of observation.  An example of such a flamboyant
poststarburst object (with $<$Balmers$>$=9.1 \AA) is shown in Figure 3. 

The emission-line galaxy population contains those galaxies which, at
the epoch of observation, are experiencing star formation at various
levels. The cutoff for inclusion into the emission line category is
the presence of [OII] 3727\thinspace\AA~emission with EW
$>$ 5 \AA, with emission line detections at a level of $>$1
$\sigma$. The variety of emission-line galaxy behaviors is shown
in Fig.~4, and includes late-type galaxies undergoing strong ([OII]
3727\thinspace\AA~EW $>$ 50 \AA) star formation at the
epoch of observation (number 519), galaxies with emission lines {\it
plus} strong Balmer absorption features (number 595), and red
absorption-line galaxies with weak [OII] emission (number 369). The
upper panel of Fig.~6 plots the distribution of [OII] emission line
strength versus galaxy color, demonstrating a clear blueing trend with
larger [OII] emission strength.\footnote{A number of galaxies in the
upper panel of Fig.~6 appear to have [OII] emission strengths above
our emission-line galaxy cutoff (5 \AA) yet are shown as
absorption-line galaxies ({\it crosses}). The [OII] detections for these
objects (mainly MMT spectra) are not significant at the 1$\sigma$
level.} A similar correlation between [OII] emission and galaxy color is also
seen for present-day spirals (Dressler \& Gunn 1982).

There are 16 emission line galaxies with [OII] emission and H$\delta$
absorption stronger than 4\thinspace\AA; these are labeled ``emi$+$H$\delta$''
in Table 1. The last class of emission-line objects, red galaxies with weak
emission ([OII] $<$ 20 \AA), includes galaxies 137, 143, 305, 369, and the
brightest cluster galaxy 375. These galaxies have spectra dominated by
absorption features yet also display weak emission. We infer that the star
formation histories of these galaxies are, for the most part, similar to those
of the rest of the absorption-line population except that they contain enough
residual gas to allow a low rate of continued star formation. Nonetheless,
because of the existence of emission lines in their spectra these objects are
classified as ``emi'' in Table 1. Weak emission in nearby red brightest
cluster, elliptical, and S0 galaxies is also often observed (e.g., Fisher,
Franx, \& Illingworth 1995; 1996).

The FMB study of CL 1358+62 classified 6 spectra as poststarburst. 
Here, our stricter poststarburst definition applies to only 2 of these
spectra.  The 4 remaining galaxies do show enhanced enhanced Balmer
absorption compared with the average for the absorption line population. 
These galaxies display average H$\delta$ line strengths of $\sim$4
\AA~compared to the 0.9 \AA~average for the 177 absorption line
galaxies, and are probably weaker examples of the poststarburst
phenomena.  The galaxies meeting our strict definition of poststarburst
have average H$\delta$ absorption line strengths of 6.3 \AA.  All five
of the galaxies classified by FMB as emission-line meet the
emission-line galaxy criteria outlined above. 

\subsection{Comparison with Low-$z$ Galaxy Spectra} 
In order to make possible connections between our cluster galaxy spectra and
those from known morphological types and evolutionary sequences we have
compared our CL 1358+62 spectra with spectra of present-day ($z$$\approx$0)
galaxies from the collections of Kennicutt (1992) and Liu \& Kennicutt (1995). 
These are compilations of the integrated spectra of normal, peculiar, and
merging galaxies enabling a valid comparison with spectra of distant galaxies
such as we observe in CL 1358+62. We measured the EW of H$\delta$, H$\gamma$,
H$\beta$ and [OII] 3727\thinspace\AA~for the 55 spectra in Kennicutt's atlas.
The Liu \& Kennicutt work tabulates these equivalent widths for their sample of
merging galaxies. 

To compare the star formation characteristics of the (non-merging) galaxies in
Kennicutt's and the CL 1358+62 galaxies we plot the H$\delta$ EW versus [OII]
3727\thinspace\AA~emission for both samples in the upper panel of Fig.~7. In
both cases only measurements with errors for H$\delta$ EW $<$5\thinspace\AA~are
used. The points for the CL 1358+62 spectra are coded according to the spectral
type definitions outlined in \S 5.1. One striking feature of the diagram is the
absence of poststarburst galaxies in the Kennicutt atlas satisfying our
criteria; all of Kennicutt's galaxies with strong H$\delta$ absorption also
show [OII] emission. In fact, the Kennicutt atlas contains just four galaxies
with large H$\delta$ absorption ($>$ 5 \AA) irrespective of [OII] emission
strength. In contrast, our collection of CL 1358+62 spectra includes a variety
of absorption-line, poststarburst, and emission-line galaxies with H$\delta$
absorption EW $>$ 5\thinspace\AA. 

Excluding the region occupied by the poststarburst galaxies, the
absorption-line and emission-line galaxy sequences in CL 1358+62 span a locus
similar to that of the nearby galaxies. This is interesting because the
Kennicutt atlas contains a wide range of normal and peculiar Hubble types from
the field and cluster environments: early-type elliptical and S0 galaxies,
spirals, irregulars, starburst and active galaxies, and HII regions. This
suggests that the galaxy population in CL 1358+62 also includes a wide variety
of spectral and morphological types, although the mix is different in
Kennicutts's sample and CL 1358+62. 

The lower panel of Fig.~7 is similar to the upper panel, except the
cluster galaxies are compared with the merging galaxy sample of Liu \&
Kennicutt ({\it solid squares}).  This comparison is interesting because
galaxy interactions have been offered as a possible mechanism leading to
the formation of poststarburst galaxies (Lavery et al.~1992; Dressler
et al.~1994; Couch et al.~1994).  In particular, Zabludoff et al.~(1996)
selected a sample of 21 E+A galaxies from the Las Campanas Redshift
Survey (LCRS; Shectman et al.~1996) and found that at least five
displayed clear tidal features, suggesting that galaxy interactions or
mergers could have played a role in their creation. 

However, the lower panel of Fig.~7 shows that poststarburst galaxies meeting
our criteria are absent from the Liu \& Kennicutt catalog. Those galaxies with
high H$\delta$ absorption ($>$4\thinspace\AA) EW also display at least moderate
[OII] emission (5\thinspace\AA$<$ [OII] EW $<$15\thinspace\AA). Liu \&
Kennicutt suggest that a number of galaxies in their sample with H$\delta$ EW
$>$6\thinspace\AA~and [OII] EW $<$15\thinspace\AA~have spectra similar to the
poststarburst (E+A) galaxies in distant clusters. However, these merging galaxy
spectra more closely resemble the emi+H$\delta$ spectral type (\S 5.1) due to
the choice of an [OII] EW cutoff of 15\thinspace\AA~. These H$\delta$-strong
galaxies with emission have been observed in other intermediate redshift
clusters (e.g.~Abraham et al.~1996) and may be in an evolutionary phase
distinct from that of the E+A galaxies. 

\section{Cluster Structure as a Function of Spectral Type} 
It has been frequently noted that early-type galaxies predominate in dense
cluster environments (Hubble \& Humason 1931; Morgan 1961; Abell 1965; Oemler
1974), and one of our prime objectives in studying CL 1358+62 is to
search for evolution in the morphology-density or morphology-radius
relation.  The van Dokkum et al.~(1997) study gives morphological
classifications for 194 galaxies in our sample that lie within the
boundaries of the mosaic of available HST images.  A future paper will
investigate in depth the relationship between these galaxy morphologies
and the spectral types presented here.  Here, we note only that there is
a clear connection between the spectral types that we define and the galaxy
morphologies.  Our galaxies with absorption-line spectra typically display
elliptical and S0 galaxy morphologies while the galaxies with strong
emission lines correspond to late-type spiral and irregular galaxies. 

\subsection{Spectral Type as a Function of Local Galaxy Density}
Here, we study the distribution of the different spectral types as a function
of local galaxy surface density and compare this with Dressler's galaxy
morphology-density relation.  We calculate surface number densities
according to the Dressler (1980) prescription: for each cluster galaxy
the ten nearest cluster galaxies are found and the distance to the tenth
object is used to compute the area enclosing all 11 objects
($\pi$$r^2_{10}$).  A local density, $\rho_{\rm
proj}$=11/$\pi$$r^2_{10}$, can then be assigned to each galaxy.  The
resulting relation between spectral type and galaxy surface density for
CL 1358+62 is shown in the upper panel of Figure 8.  There are 46 or 47
galaxies in each of the bins in Fig.~8. 

From Fig.~8 it is clear that the distributions of spectral types do vary
as a function of local galaxy density.  A Kolmogorov-Smirnov (KS) test
indicates that the distribution of absorption-line and emission-line
galaxies with galaxy density were not drawn from the same parent
population with 99.84\% confidence.  In our highest density bin ($>$71
galaxies Mpc$^{-2}$ with average $\rho_{\rm proj}$=168 galaxies
Mpc$^{-2}$) 87$\pm$14\% have pure absorption line spectra, 9$\pm$4\%
have emission line spectra and 4$\pm$3\% have poststarburst spectra. 
In the highest density regions studied by Dressler (1980) in rich
clusters at low redshift ($\rho_{\rm proj}$=100 gals Mpc$^{-2}$) he
found 88\% S0+E galaxies and 12\% S+Irr galaxies. 

In the lowest density regions of CL 1358+62 ($\rho_{\rm proj}$$<$13 gals
Mpc$^{-2}$ with average $\rho_{\rm proj}$=8 gals Mpc$^{-2}$)
54$\pm$15\% have pure absorption line spectra and 46$\pm$14\% have
emission line spectra.  This census may be compared with a similar low
density region (average $\rho_{\rm proj}$=1.3 gals Mpc$^{-2}$) from
Dressler's cluster sample which contained 43\% E+S0 galaxies and 57\%
S+Irr galaxies.  These comparisons suggest that the spectral
type-density relation in CL 1358+62 resembles in broad outline the
morphology-density relation in Dressler's clusters at $z$$\sim$0.04. 

The fraction of poststarburst galaxies may peak at intermediate galaxy
surface density ($\rho_{\rm proj}$$\approx$35 gal Mpc$^{-2}$).  No
poststarburst galaxies are found in our lowest density bin.  However,
there are only 11 poststarburst galaxies in our sample so firm
conclusions regarding their distribution await additional studies of
intermediate redshift clusters. 
	
\subsection{Spectral Type as a Function of Clustercentric Radius} 
The lower panel of Figure 8 is a plot of the relative number fractions of
the spectral types versus projected radius from the BCG.  The
absorption-line galaxies, the dominant population, define the cluster
structure and show a smooth decrease in number fraction with increasing
clustercentric radius.  The fraction of emission-line galaxies increases
steadily with clustercentric radius.  The upper and lower panels of
Figure 8 are qualitatively similar.  A KS test indicates that at the
99.94\% confidence level, the distribution of absorption-line and
emission-line galaxies shown in the lower panel of Figure 8
are not drawn from the same parent population.  At 97.52\%
confidence the absorption-line and H$\delta$+emission line galaxies
are not drawn from the same parent population. 

Within a projected radius of $r_{\rm p}$$<$0.4 Mpc 86$\pm$17\% of the
galaxies have absorption line spectra, 8$\pm$6\% have poststarburst
spectra, and 8$\pm$6\% have emission-line spectra.  This mix is
consistent with the galaxy populations found in other cluster cores at
intermediate redshift (e.g., Abraham et al.~1996).  In the outermost
radial bin (at $r_{\rm p}$=1.7 Mpc) 59$\pm$11\% of the galaxies have
pure absorption-line spectra and 41$\pm$9\% have emission-line spectra. 
Of the latter group, 70\% display vigorous star formation ([OII] EW $>$
20\thinspace\AA).  There are no poststarburst galaxies in the outermost
radial bin. 

\subsection{Projected Distribution as a Function of Spectral Type} 
The distributions of the various spectral types on the plane of the sky are
shown in Figure 9. For the emi+H$\delta$, emission-line and poststarburst
galaxies, the point size indicates the strength of H$\delta$ absorption, [OII]
emission, and Balmer line absorption, respectively. The absorption-line
galaxies, as the dominant population, define the cluster structure while the
poststarburst and emission-line galaxies are less clustered. A two-dimensional
KS test indicates that the spatial distribution of the spectroscopically active
galaxies could not have been drawn from the same parent sample as the
absorption-line galaxies with 99.28\% confidence.

The spatial distribution of emission-line galaxies appears to be
nonuniform: 31 emission-line galaxies lie southwest of a line drawn
through the origin from SE to NW, while 12 emission-line galaxies lie to the
northeast of the line. We note that similar nonuniformities in emission-line
galaxy spatial distributions have been seen in other intermediate redshift
clusters (e.g., Abraham et al.~1996) and were used to argue for a preferred
directions of subclustering or infall. However, simulations indicate that the
emission-line galaxy distribution in CL 1358+62 is inconsistent with a
symmetric distribution at only 82\% confidence.

The poststarburst galaxy distribution in Fig.~9 appears to be aligned linearly
along a roughly North-South direction. We note that the substructuring we find
in CL 1358+62 and that found by the CNOC group (see \S 7) are also located
along a N-S axis. However, we find that the apparent departure from a symmetric
distribution is not statistically significant using a two-dimensional KS test.

\subsection{Velocity Distributions of the Spectral Types} 

We use the prescriptions of Danese, De Zotti, \& di Tullio (1980) to
calculate the means and dispersions for the entire sample and for the
different galaxy spectral types.  We find a mean $cz$=98425$\pm$90 km
s$^{-1}$ ($z$= 0.32831$\pm$0.00030) and a velocity dispersion of
1027$^{+51}_{-45}$ km s$^{-1}$ for the entire sample of 232 galaxies. 
Figure 10 presents the velocity histograms for the absorption-line,
poststarburst, emission-line, and emi$+$H$\delta$ galaxies.  It appears
that the absorption-line galaxies have a lower dispersion than the
poststarburst and emission-line galaxies. 

The mean velocities and dispersions for each of the spectral types are
summarized in Table 3.  We also give the kinematic parameters for the 23
emission-line galaxies with weak [OII] emission (5 \AA~$<$[OII] EW$<$20
\AA) and for the 21 with strong [OII] emission([OII] EW$>$ 20 \AA.  None
of the subgroupings of emission-line galaxies (the full sample of 44
emission-line galaxies, weak emission-line, strong emission-line, and
the emi$+$H$\delta$) have mean velocities which differ by more than 1
$\sigma$ from the mean velocity of the dominant absorption-line
population. 

The mean $cz$ of the poststarburst galaxies differs from that of the
absorption line galaxies by 1 $\sigma$ and the dispersion differs by
$\sim$2 $\sigma$.  A KS test indicates that at 90\% confidence the two
velocity distributions are not drawn from the same parent population. 
We note that the velocity dispersion of freely infalling galaxies is
expected to be 2$^{1/2}$ times that of the virialized cluster core
population.  The ratio of the velocity dispersions of the spirals and
ellipticals in both the Virgo (Huchra 1985) and Coma (Colless \& Dunn
1996) clusters are of this order.  Although the statistics are weak at
present, the ratio of the velocity dispersions of the poststarburst and
absorption-line galaxies in CL 1358+62 are also $\sim$2$^{1/2}$. 

The evidence for dynamical differences as a function of spectral type
is strongest when we combine the poststarburst and emission line
galaxies in a single ``active" population.  For galaxies with velocities
within $\pm$0.6 $\sigma_{\rm cl}$ of the cluster mean the active
fraction is 16$\pm$4\%, rising to 32$\pm$7\% in the tails of the
velocity distribution ($\sim$2.0 $\sigma_{\rm cl}$ from the mean). 

\section{Subclustering} 

The many successful searches for velocity substructure in well-sampled
galaxy clusters suggest that clusters continue to accrete galaxies to
the present day (e.g., Geller \& Beers 1982; Dressler \& Shectman 1988;
Fitchett \& Webster 1987; Zabludoff, Franx, \& Geller 1993; Colless \&
Dunn 1996).  The manifestations of cluster substructure are varied and a
multitude of tests have been devised to measure various types of
substructures.  These include counting nearest neighbors that satisfy
velocity and dispersion criteria (Dressler \& Schectman 1988), the KS
two-sample test (Colless \& Dunn 1996), searching for asymmetries in the
shape of the velocity distribution (Zabludoff, Franx, \& Geller 1993),
and mixture-modeling algorithms (McLachlan \& Basford 1988).  No single
technique is superior in all situations; their sensitivities to
particular aspects of the observed velocity distribution vary.  The
various techniques differently weight the tails of the velocity
distribution, regions of high or low galaxy density, or structures
resolved in velocity space, but superposed spatially, such as cluster
cores. 

We too have used a variety of methods to search for subclustering within
CL 1358+62.  Figure 11 plots redshift against projected radius.  The
cluster mean velocity as well as the redshift boundaries used to define
cluster membership are shown (see \S 4).  We see an excess of galaxies
with redshifts lower than the cluster mean within a radius of
40\arcsec~(0.23 Mpc). 

Following the technique outlined by Zabludoff, Franx, \& Geller (1993)
we have decomposed the full cluster velocity distribution into a sum of
orthogonal Gauss-Hermite functions to quantify the asymmetric
third-order ($h_3$) and symmetric fourth-order ($h_4$) terms of the
distribution.  This technique indicates that the velocity distribution
of CL 1358+62 departs from a Gaussian with both asymmetric and symmetric
distortions.  We measure an $h_3$ of $-$0.088$\pm$0.045 for the velocity
distribution of all 232 cluster members.  This result is significant at
the 94.6\% level.  This asymmetric deviation is in the sense that CL
1358+62 has a tail of galaxies at redshifts lower than the cluster mean. 
The $h_4$ term for the velocity distribution of CL 1358+62 is
$-$0.087$\pm$0.042, which indicates that symmetric deviations are also
present in the sense that the velocity distribution is broader than a
Gaussian. This $h_4$ term is significant at the 97.8\% confidence
level. 

We have calculated the Dressler \& Shectman (1988) substructure
statistic $\delta$ for each cluster member in CL 1358+62.  The $\delta$
parameter is a measure of the deviation of any one galaxy from the local
velocity mean and dispersion.  In Figure 12 we plot $\delta$ as a
function of projected position within the cluster.  The results indicate
that significant kinematic irregularities are present along the cluster
center line-of-sight and towards the northwest corner. 

The significance of these deviations has been calibrated with Monte
Carlo simulations.  A total of 1000 realizations were calculated in
which the velocities of the 232 cluster members were randomly reassigned
while fixing the galaxy positions.  The cumulative deviation $\Delta$,
the sum of the $\delta$'s for the individual cluster members, is 315.6. 
According to our simulations, this cumulative deviation is significant
at the 96\% confidence limit.  If the velocity distribution of CL
1358+62 were close to Gaussian and the local variations were random,
then $\Delta$ would be of the same order as the total number of cluster
members. 

In Figure 13 we plot the Dressler-Shectman statistic as a function of
velocity and each of two projected coordinates ($x$ and $y$). We see
that the dominant kinematic substructure is a group of 20 galaxies with
velocities 95000$<$$cz$$<$97000 projected close to the cluster center
(projected radius $r_{\rm p}$$<$95\arcsec~centered at
$x$=$-$27\arcsec~$y$=$-$8\arcsec). This is the same foreground group
mentioned in connection with Fig.~11 that produces the low velocity tail
found by the Gauss-Hermite analysis. The mean velocity of this group is
96272$\pm$119 km s$^{-1}$ with $\sigma$=391$^{+84}_{-53}$. We find a
second subclump composed of 10 galaxies with velocities
99000$<$$cz$$<$101000, centered at $x$=77\arcsec~$y$=114\arcsec with
$r_{\rm p}$$<$60\arcsec. The mean velocity of this group is
99771$\pm$136 km s$^{-1}$ and $\sigma$=314$^{+112}_{-56}$. 

Figure 14 plots contours of galaxy surface density in radial velocity
intervals of 2000 km s$^{-1}$. The foreground subclump discussed above
appears in the first velocity bin as a dense concentration of galaxies
towards the cluster center. The middle velocity bin includes the bulk
of the cluster members that are roughly spherically distributed around
the BCG at $x$=0 $y$=0. The background velocity bin shows a relatively
empty cluster core as well as two areas of higher concentration to the
north (the second subclump identified above) and south of the cluster
center. 

The foreground subclump (at a projected radius of 0.2 Mpc in a region with an
average surface density of 125 galaxies Mpc$^{-2}$) consists of 80\%
absorption-line, 15\% poststarburst, and 5\% emission-line galaxies. This mix
does not deviate significantly from the the cluster average at this projected
radius and surface density. All 10 of the galaxies in the background subclump
north of the cluster center have pure absorption-line spectra. This subclump
has a projected radius of 0.8 Mpc and an average galaxy surface density of 46
galaxies Mpc$^{-2}$. At this projected radius and surface density the cluster
as a whole contains $\sim$80\% absorption line galaxies. Therefore, the
subclumps do not appear to contain a peculiar mix of spectral types.
Furthermore, the spectroscopically active galaxies do not have an excess or
deficit of nearby companions. 

The redshift distribution of CL 1358+62 has been studied previously by the CNOC
group (Carlberg et al.~1996; Carlberg, Yee, \& Ellingson 1997) who have amassed
a catalog of 164 cluster galaxy redshifts. The redshift boundaries used for
determining cluster membership are very similar for both studies. The CNOC
field limits are 2\arcmin~narrower in R.A.~than ours and extend to the same
Declination to the north. However, the CNOC field extends 4\arcmin~further to
the south and here the CNOC group detects a concentration of galaxies with $cz$ 
differing by 1000 km s$^{-1}$ from the cluster center's $cz$. This feature
leads the CNOC group to classify CL 1358+62 as a binary cluster. 

\section{Summary and Conclusions} 

Our investigation of CL 1358+62 is based on a catalog of 232 cluster galaxy
spectra. We combine the velocity, spatial, and spectral information from our
catalog to study the cluster structure. Similar to other rich clusters, the
dominant galaxy population in CL 1358+62 is composed of red, absorption-line
galaxies with passively evolving stellar populations that formed the majority
of their stars long ($>$ 5 Gyr) before the epoch of observation. The
absorption-line galaxies make up 76\% of the total cluster population, rising
to $\sim$90\% of the galaxies with velocities within 0.6 $\sigma_{\rm cl}$ of
the cluster mean and located within the central $r_{\rm p}$$<$0.4 Mpc. 

Our sample concentrates on the central main body of the cluster 
($r_{\rm p}$$<$2 Mpc). We note that the CNOC project had previously found
evidence for a second concentration south of the cluster center, just outside
our field limits (Carlberg et al.~1996, Carlberg, Yee, \& Ellingson 1997).
Significant evidence for substructure in the central part of the cluster is
seen. We find that the distribution of line-of-sight velocities for CL 1358+62
departs significantly from a Gaussian. Two compact (with projected radii
$r_{\rm p}$$<$90\arcsec~and $\sigma$$\leq$400 km s$^{-1}$) subclumps are found
in the cluster, one with a mean $z$ higher than that of the cluster as a whole,
and one with a lower mean $z$. This subclustering implies that CL 1358+62 has
not yet reached equilibrium and is still in the process of virialization and
accretion of new members. The galaxies in these subgroups have primarily
absorption-line spectra, indicating that accretion into groups may play an
important role in the spectral evolution of galaxies. 

The relations between spectral type and radius/galaxy density in CL 1358+62  
are qualitatively similar to the morphology-radius or morphology-density 
relations seen in nearby clusters (i.e., Dressler 1980; Whitmore, Gilmore, \&  
Jones 1993). That the galaxy mix in CL 1358+62 has reached proportions similar
in magnitude and distribution to those seen in nearby rich clusters is
indicative of its advanced state of evolution. Similar results have been seen
in the cores of other intermediate redshift clusters (Couch et al.~1997;
Dressler et al.~1997) which are dynamically evolved and centrally concentrated.

The gradients in the spectral mix as a function of radius and local density
indicate that the mean ages of the galaxy stellar populations decrease with
increasing distance from the cluster center. This is evidenced by the increase
in emission-line galaxy fraction with increasing clustercentric radius. These
results are consistent with the idea that the galaxy evolution in CL 1358+62
is shaped by the accretion of infalling field galaxies whose star formation
rates are diminished upon cluster entry. Similar conclusions were reached by
Abraham et al.~(1996) in their study of the intermediate redshift ($z$=0.228)
cluster Abell 2390.

Spectrally active galaxies make up 23$\pm$3\% of the cluster population; 4/5 of
these have emission-line spectra indicating continuing star formation while the
remainder have poststarburst (``E+A'') spectra. This fraction of spectrally
active galaxies is consistent with the elevated proportion (compared to
low-$z$) observed for samples of distant clusters at similar redshifts
(Dressler \& Gunn 1983; Couch \& Sharples 1987; Abraham et al.~1996). For CL
1358+62, the fraction of spectroscopically active galaxies increases in the
tails of the velocity distribution, lending further support to the hypothesis
that galaxies with current or recent signs of star formation have recently been
accreted. However, the high relative velocities could also be explained if the
spectrally active galaxies were preferentially located in radial orbits. 

Poststarburst galaxies make up 5\% of the CL 1358+62 cluster population. This
can be compared to the Las Campanas Redshift Survey (LCRS), which contained
just 0.2\% ``E+A'' galaxies (Zabludoff et al.~1996). This sample is typical of
the nearby field (0.05$<$$z$$<$0.13). However, the definition Zabludoff et
al.~applied to select poststarburst galaxies was stricter than the one used
here. If our definition is used, 1\% of the LCRS sample are poststarburst
galaxies (Zabludoff, private communication). The origin of these field
poststarburst galaxies is still an open question although galaxy-galaxy
interactions have been offered as a possibility (Zabludoff et al.~1996).

We have addressed the question of whether the poststarburst galaxy spectra in
CL 1358+62 resemble those found for major mergers by making a comparison
between our ``E+A'' spectra and spectra of nearby merger remnants from
Kennicutt (1992) and Liu \& Kennicutt (1995). We find that none of the major
merger spectra satisfy our poststarburst selection criteria. All of the nearby
merger remnants have detectable [OII]\thinspace3727\AA~emission with equivalent
widths larger than 5\AA. This implies that the CL 1358+62 ``E+A'' galaxies are
either not formed by major mergers, or that a process operates that removes
the gas from the mergers very quickly after the merger event. Interactions with
and/or stripping by the intracluster medium (Gunn \& Gott 1972; Bothun \&
Dressler 1986) may be an effective mechanism for removing the remaining gas in
merging cluster galaxies. We note that galaxy-galaxy interactions that are
less severe than the major mergers of Liu \& Kennicutt have also been offered
as a mechanism for truncating star formation in accreting galaxies (Moore et
al.~1996).

If larger samples confirm our tentative observation that poststarburst galaxies
are absent in the outermost, lowest density regions of clusters, we might
conclude that an infalling galaxy must penetrate to regions where galaxy-galaxy
interactions, stripping from the intracluster medium, or interaction with the
cluster potential become strong enough to drive their evolution. The 
H$\delta$-strong galaxies with [OII] emission might then represent an
intermediate state in a single evolutionary sequence from infalling
spiral/irregular to poststarburst galaxy. If so, then the emi+H$\delta$
galaxies must abruptly exhaust their gas supply and terminate star formation
while the Balmer absorption line strengths are still large. We are currently
comparing the morphologies of the emi+H$\delta$ and poststarburst galaxies in
CL 1358+62 (using the HST images, \S 2) in order to address this issue. We note
that Barger et al.~(1996) have constructed models that suggest similar
evolutionary sequences.

The catalog that we present for CL 1358+62 at $z$=0.328 is the largest
available for a cluster at $z$$>$0.1. In a companion study, van Dokkum et
al.~(1997) investigated the color-magnitude relation for CL 1358+62 using a
wide field mosaic of multi-color HST WFPC2 images. This combination of a
magnitude-limited spectroscopic survey and high resolution HST imaging has
allowed us to study CL 1358+62, an object with a lookback time $\sim$1/3 the
present age of the Universe, at a level of detail heretofore only attained for
a tiny handful of nearby clusters. Our overall goal is to elucidate the key
mechanisms governing galaxy evolution by obtaining HST mosaics and ground-based
spectra for rich clusters at increasing redshift. 

Although our program is initially focussed on cluster galaxies, we also
serendipitously obtain spectra and images of the foreground and background
galaxies along the cluster line-of-sight (e.g.~Franx et al.~1997). We are 
therefore able to study the stellar populations of field objects with the same
level of detail as our cluster galaxies. The spectral and WFPC2-derived
morphological characteristics of these galaxies are currently being examined
and results will be presented in forthcoming works. 

\acknowledgments
MF thanks the Center for Astrophysics for its hospitality during several
visits. We gratefully acknowledge support from a grant of the University
of Groningen.

This research has made use of the NASA/IPAC Extragalactic Database (NED) which
is operated by the Jet Propulsion Laboratory, Caltech, under contract with the
National Aeronautics and Space Administration.

\newpage

\figcaption[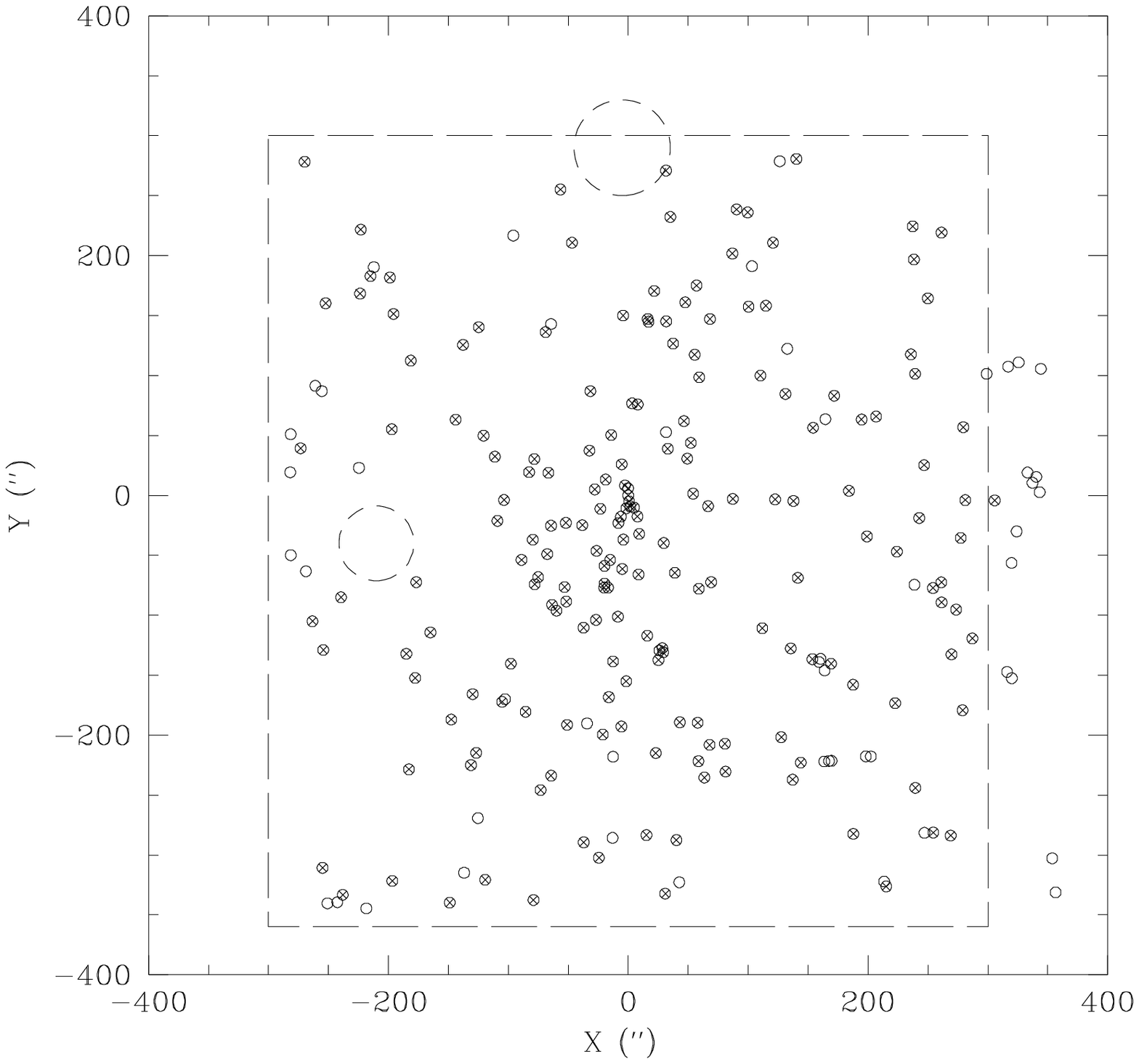]
{Spatial distribution for all galaxies ({\it open circles}) in our photometric
catalog with R magnitudes brighter than 21. Galaxies for which redshifts were
measured are marked by {\it crosses}. The two regions enclosed by the {\it
large dotted circles} indicate areas where bright stars prevented data from
being collected. The rectangular region outlined by the dotted lines delineates
the area where we have concentrated our spectroscopy.}

\figcaption[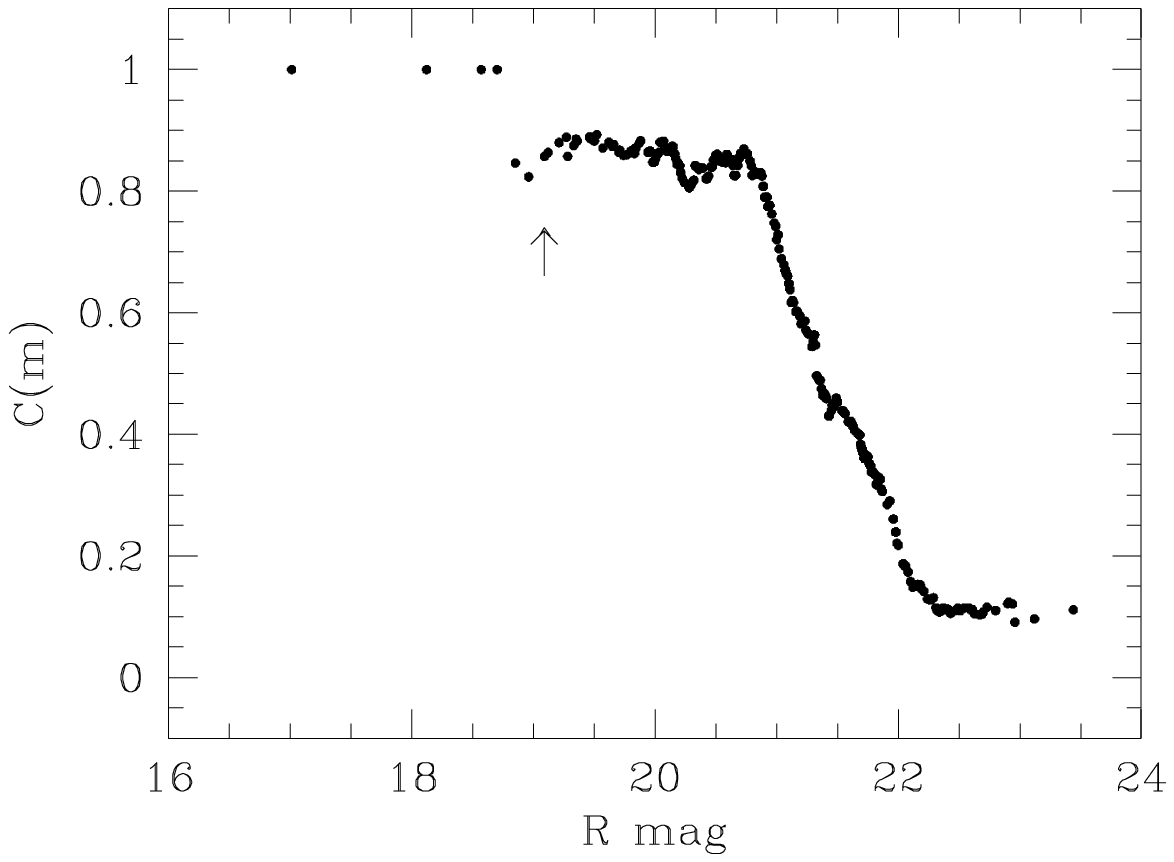]
{Completeness of the redshift sample as a function of magnitude. Each point
gives the ratio of the number of galaxies for which redshifts were measured to
the total number of possible galaxies. The completeness is shown for each 
galaxy in running bins of $\pm$0.50 mag for galaxies brighter than $R$=20.0
mag, and bins of $\pm$0.25 mag for galaxies fainter than $R$=20.0 mag. The
position of the brightest cluster galaxy is marked with an arrow.}

\figcaption[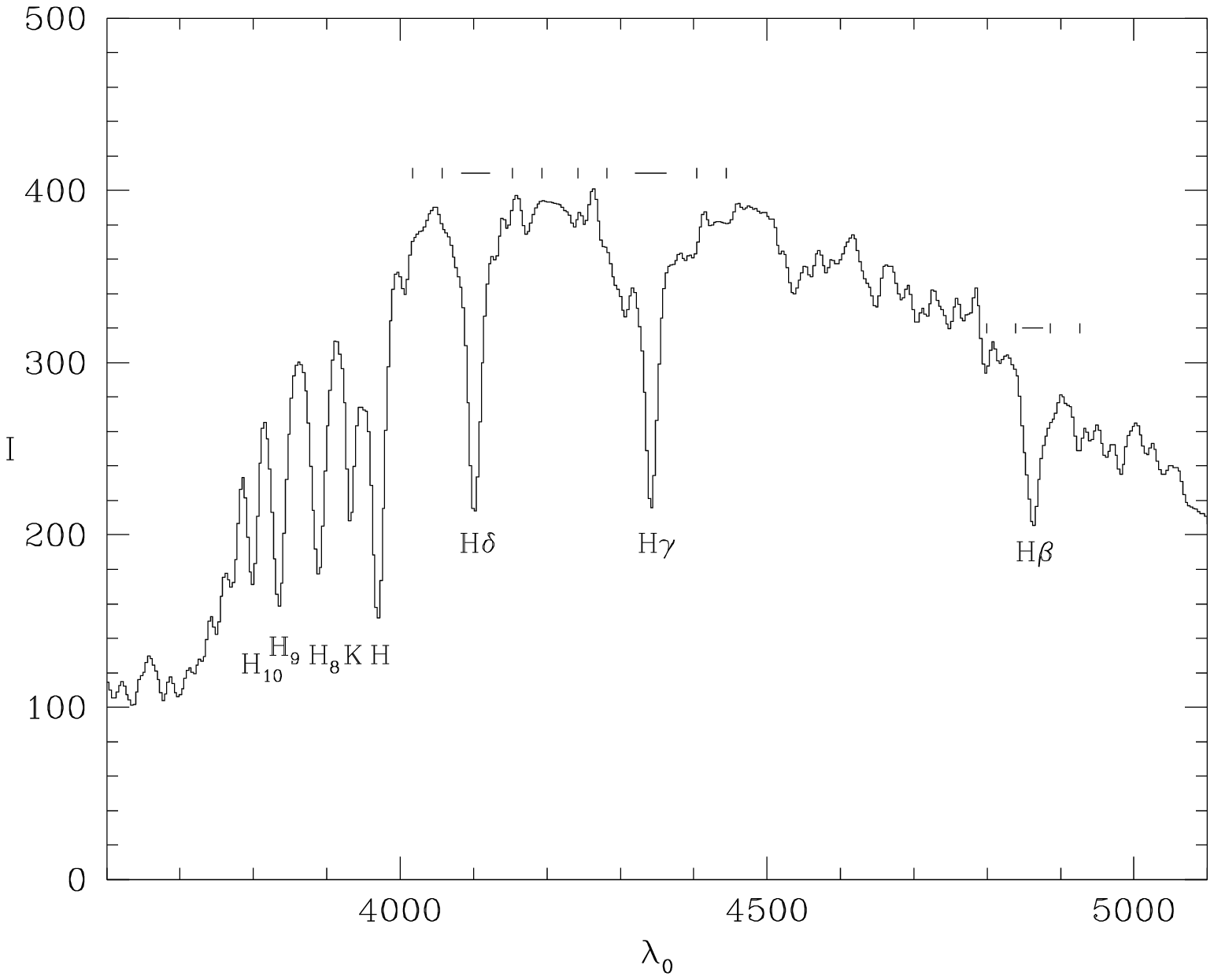]
{Deredshifted spectrum of the poststarburst cluster galaxy 243 with the Balmer
line strength indice definitions superimposed. The horizontal lines delineate
the regions covered by the feature bandpasses and the pairs of vertical lines
on each side of the indice show the regions covered by the associated continuum
sidebands.}

\figcaption[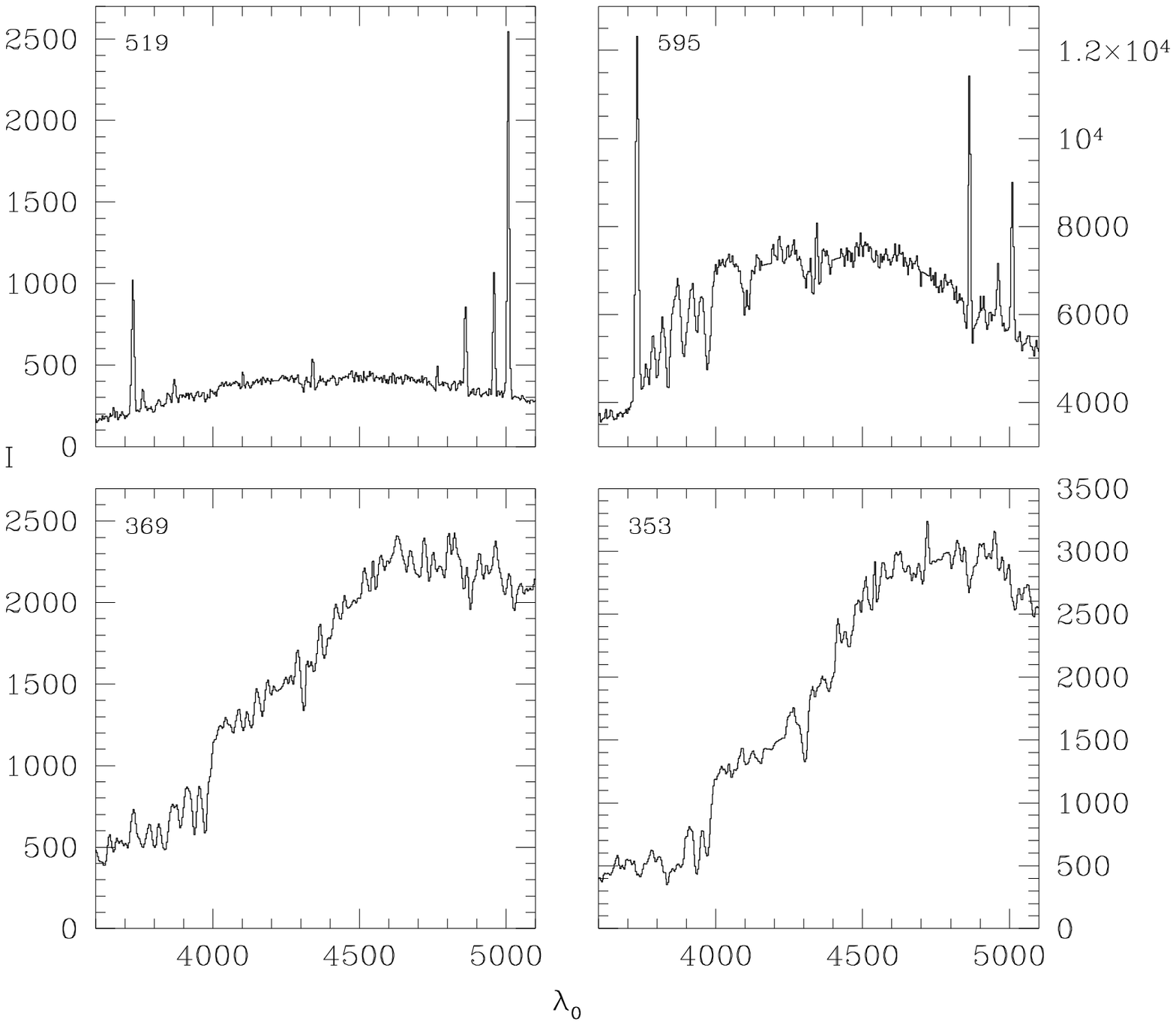]
{Examples of spectra for the various galaxy spectral types. The object
identifier is given for each of the deredshifted galaxies. The top left panel
displays the spectrum of a strong emission line galaxy (519) with very blue
colors and a host of emission lines: [OII] 3727\thinspace\AA, [NeIII]
3869\thinspace\AA, H$\delta$, H$\gamma$, [OIII] 4959\thinspace\AA, H$\beta$,
[OIII] 5007\thinspace\AA, and H$\alpha$. The top right spectrum is for a 
late-type galaxy (595) in CL 1358+62 showing both emission and strong balmer
absorption features. The lower left panel displays the spectrum of an
absorption-line galaxy (369) with weak [OII] 3727\thinspace\AA~emission and the
lower right panel shows the spectrum from one of the brighter absorption-line
galaxies (353) in the cluster with no emission features.}

\figcaption[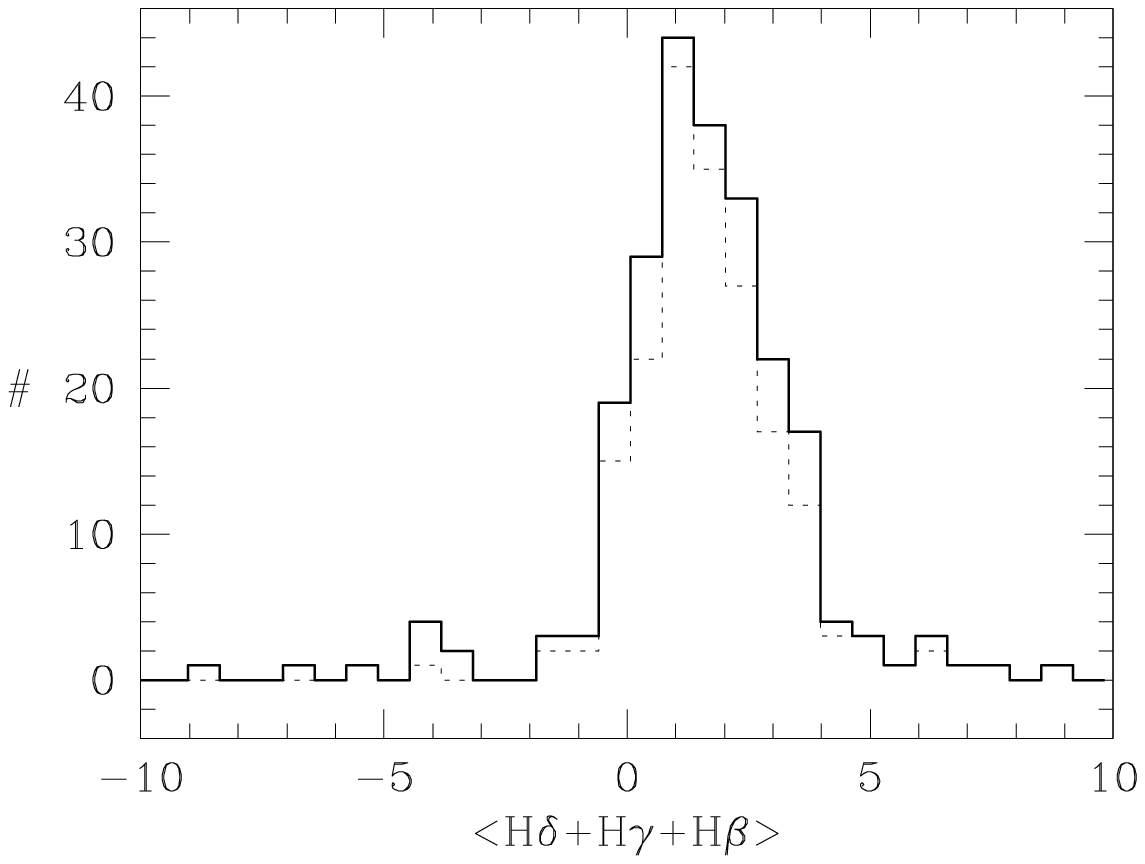]
{Distribution of the average Balmer line strength,
(H$\delta$+H$\gamma$+H$\beta$)/3, for the full sample of cluster galaxies
({\it bold solid line}) and for those without measureable OII
3727\thinspace\AA~emission ({\it dotted line}). The distributions display a
tail of galaxies with high Balmer line strengths (average$>$4.0\thinspace\AA)
which we use to identify the poststarburst population.}

\figcaption[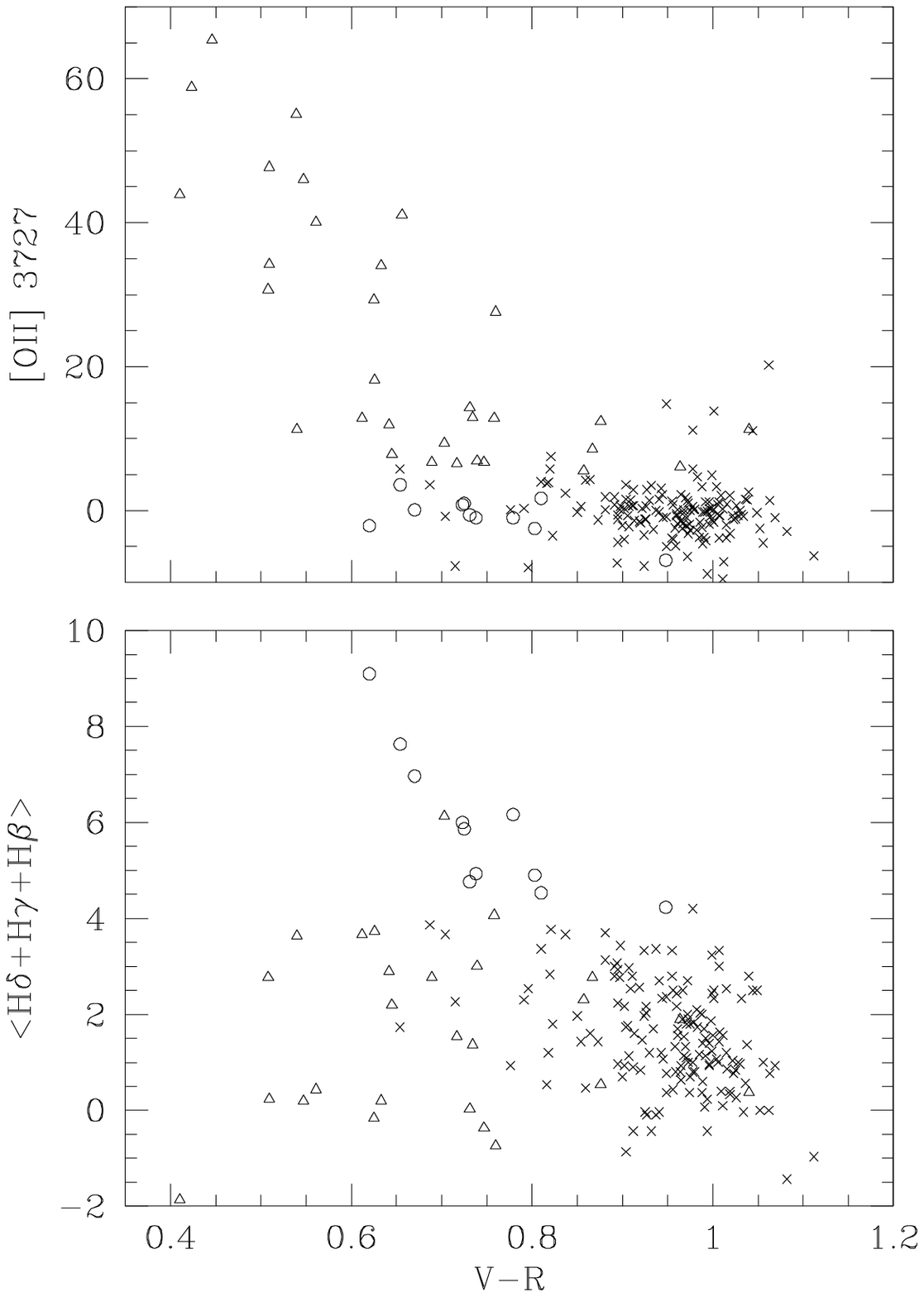]
{Balmer absorption and emission line strengths as a function of galaxy color.  
{\it Crosses} are used to represent the normal absorption-line dominated 
red galaxy population, {\it triangles} are shown for all the galaxies with the
[OII] 3727\thinspace\AA~emission line in their spectra, and {\it circles} are
used for the galaxies displaying poststarburst (no emission lines and strong
Balmer absorption features) spectra. Note the clear separations of these
different spectral types in these diagrams.}

\figcaption[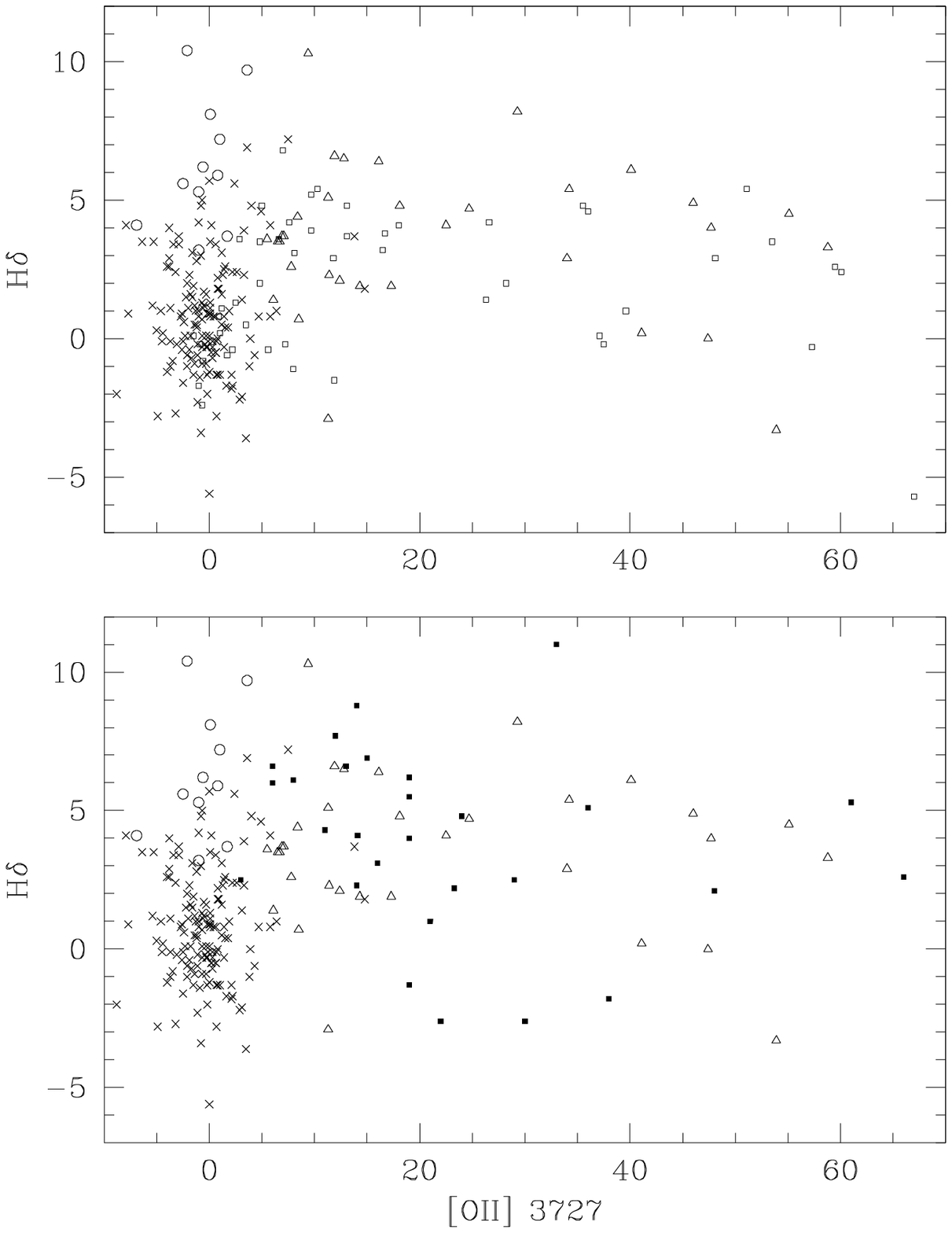]
{H$\delta$ line strength plotted versus [OII] 3727\thinspace\AA~emission line
strength. {\it Crosses} denote the absorption-line galaxies in CL 1358+62,
{\it triangles} are shown for the emission-line galaxies, and {\it circles}
represent the poststarburst galaxies. In the top panel {\it open squares} are
used for measurements from Kennicutt's atlas of nearby galaxy spectra and the
{\it filled squares} in the bottom panel shows results for nearby merging
galaxies from Liu \& Kennicutt.}
 
\figcaption[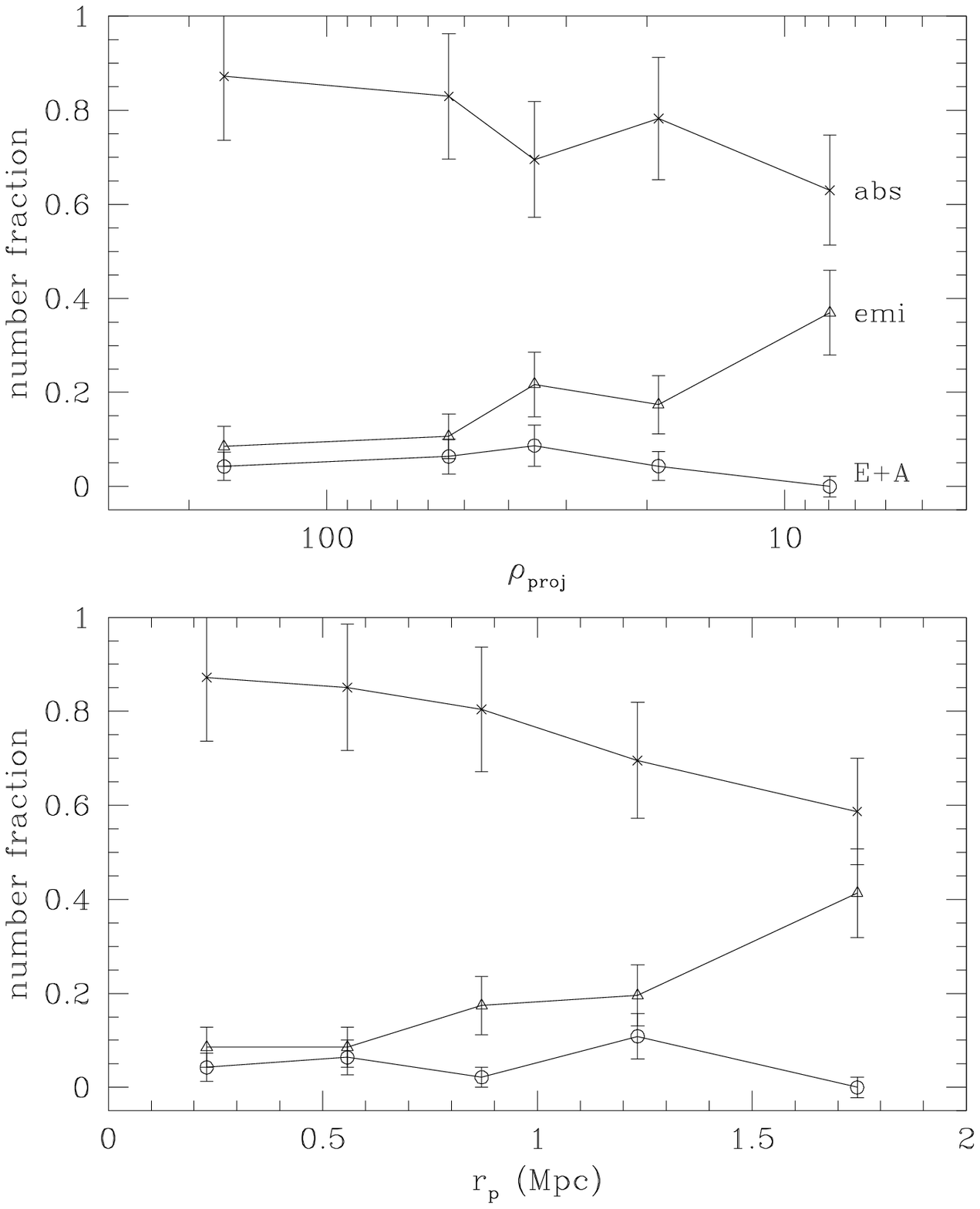]
{Number fractions of the various spectral types versus projected galaxy surface
number density and clustercentric radius. {\it Crosses} are used to represent
the normal absorption-line dominated galaxy population ({\bf abs}), {\it
triangles} are shown for all the galaxies with the [OII]
3727\thinspace\AA~emission line in their spectra ({\bf emi}), and {\it circles}
are used for the galaxies displaying poststarburst spectra ({\bf E+A}). Note
the absence of poststarburst galaxies in the low density, outermost radial
regions.}

\figcaption[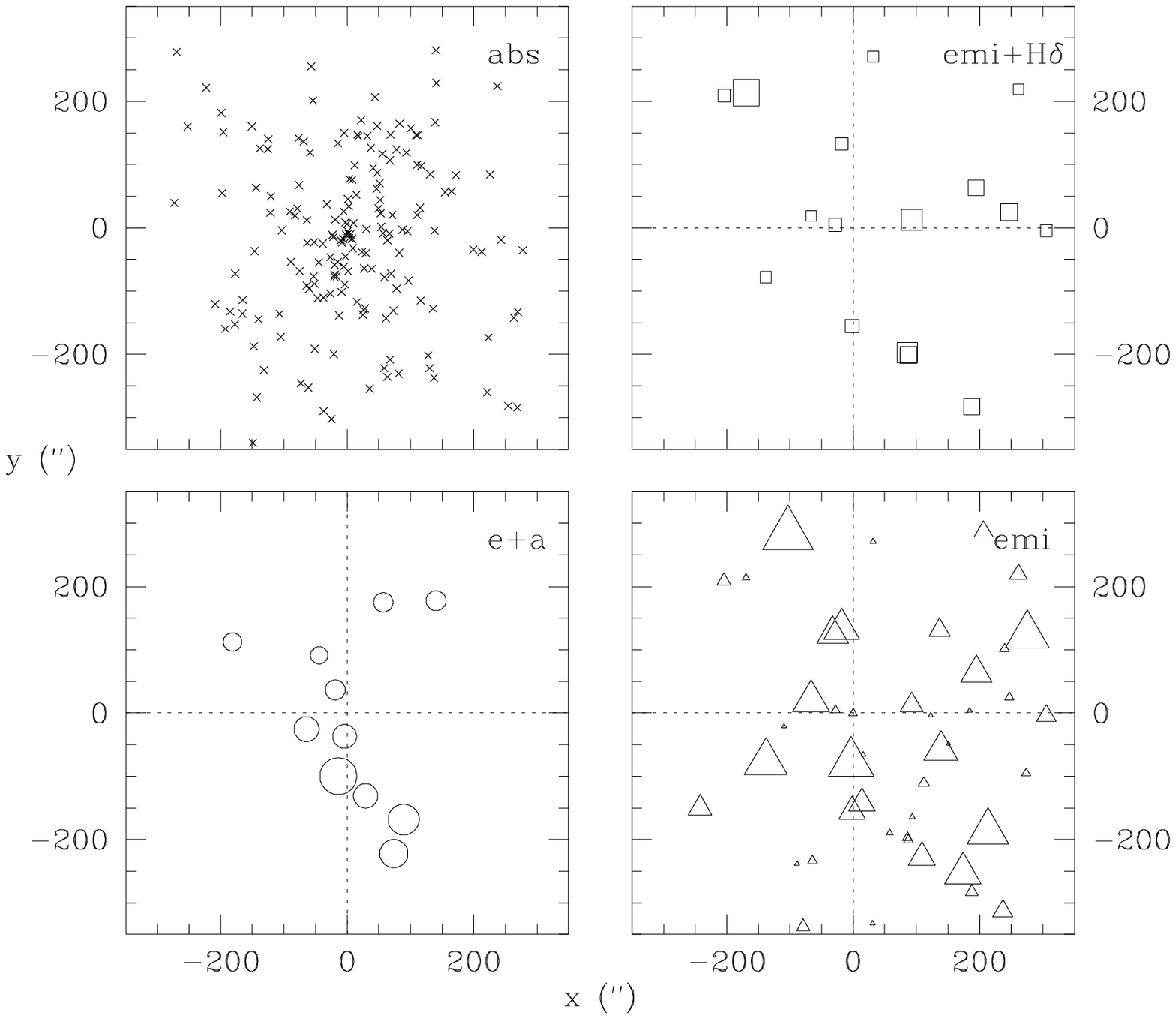]
{Spatial distributions of galaxies in CL 1358+62 as a function of spectral
type. {\it Crosses} are used to represent the normal absorption-line dominated
galaxy population ({\bf abs}), {\it squares} denote emission line galaxies
which also display strong ($>$4 \AA) H$\delta$ absorption ({\bf
emi+H$\delta$}), {\it triangles} are shown for all the galaxies with emission
lines in their spectra ({\bf emi}), and {\it circles} are used for the galaxies
displaying poststarburst (no emission lines and strong Balmer absorption
features) spectra ({\bf e+a}). The symbol sizes for the emi+H$\delta$, emi,
and E+A galaxies are scaled linearly with the strength of the H$\delta$
feature, [OII] 3727\thinspace\AA~emission line, and Balmer lines
($<$H$\delta$+H$\gamma$+H$\beta$$>$), respectively.}

\figcaption[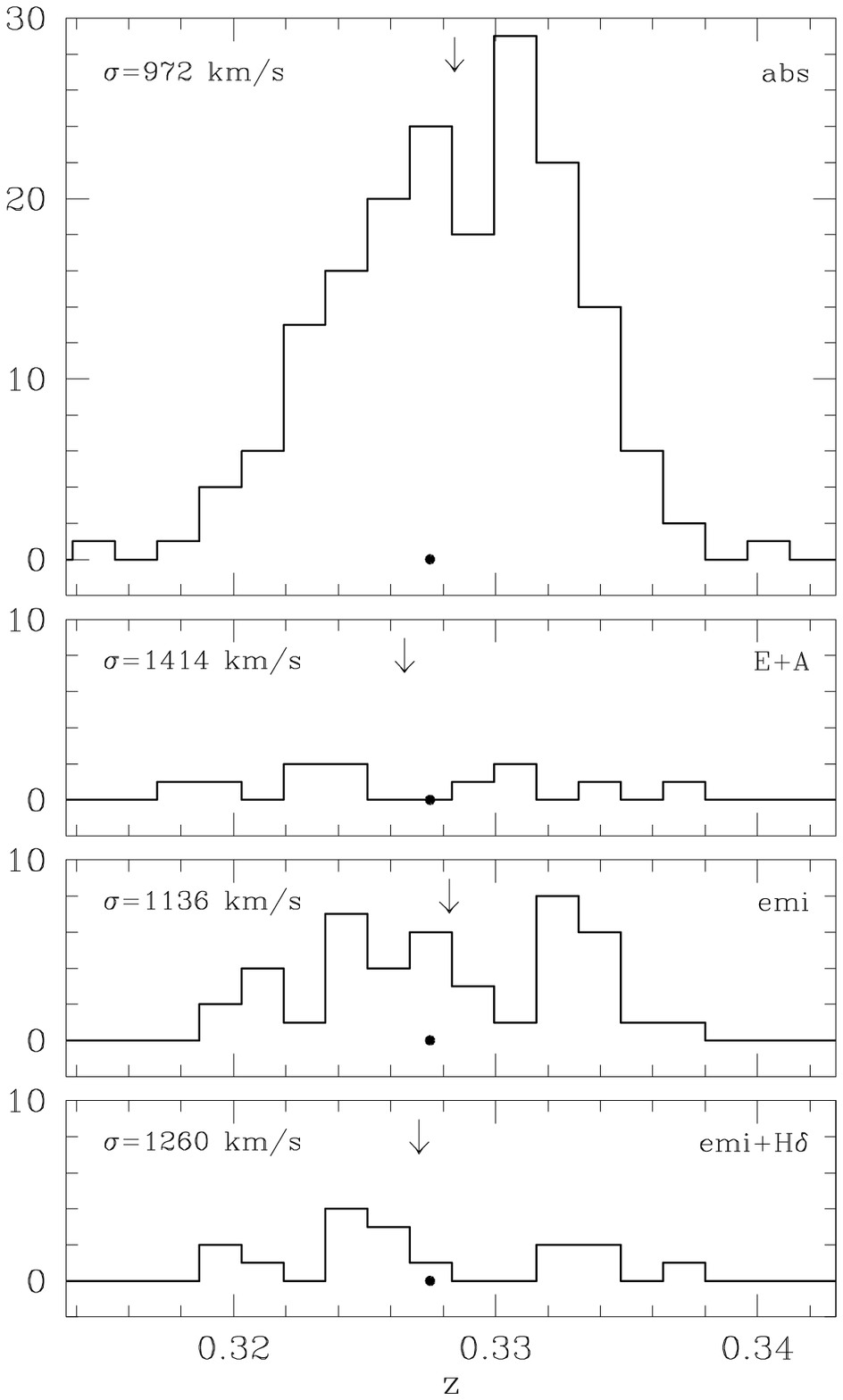]
{Velocity distributions as a function of spectral type. The absorption-line
galaxies ({\bf abs}) display a velocity distribution which is not as broad as
that shown by the emission line galaxies ({\bf emi}) and the emission-line
galaxies with H$\delta$ absorption ({\bf emi+H$\delta$}). The poststarburst
galaxies ({\bf E+A}) are uniformly distributed in redshift throughout the
cluster with large dispersion. The {\it solid dot} in each panel denotes the
redshift of the brightest cluster galaxy. The downward pointing arrow shows the
mean velocity for each distribution and the dispersion is also given.}

\figcaption[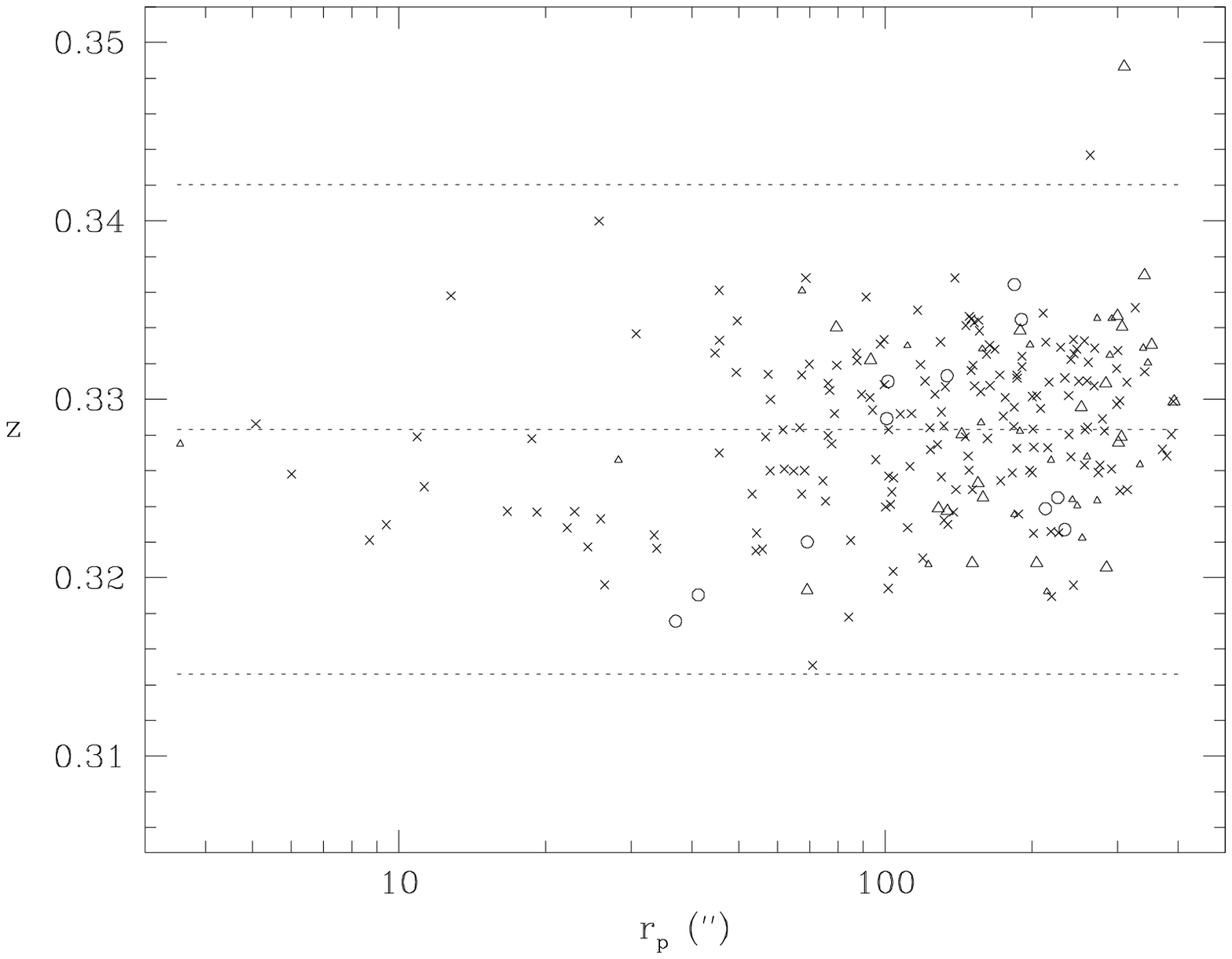]
{Redshift-projected radius distribution for the various spectral types. {\it
Crosses} are used to represent the absorption-line galaxy population 
and {\it circles} are used for the galaxies displaying poststarburst spectra.
Strong emission line galaxies ([OII] 3727\thinspace\AA$>$20) are displayed as
{\it large triangles} and weak emission line galaxies are given {\it small
triangles}. The central dotted line is located at the mean redshift of the 
cluster and the upper and lower dotted lines denote the 4 $\sigma_{\rm cl}$
boundaries used for the definition of cluster membership. The brightest cluster
galaxy is placed on the figure as the innermost r$_{\rm p}$ point (note that
the BCG displays weak emission).}

\figcaption[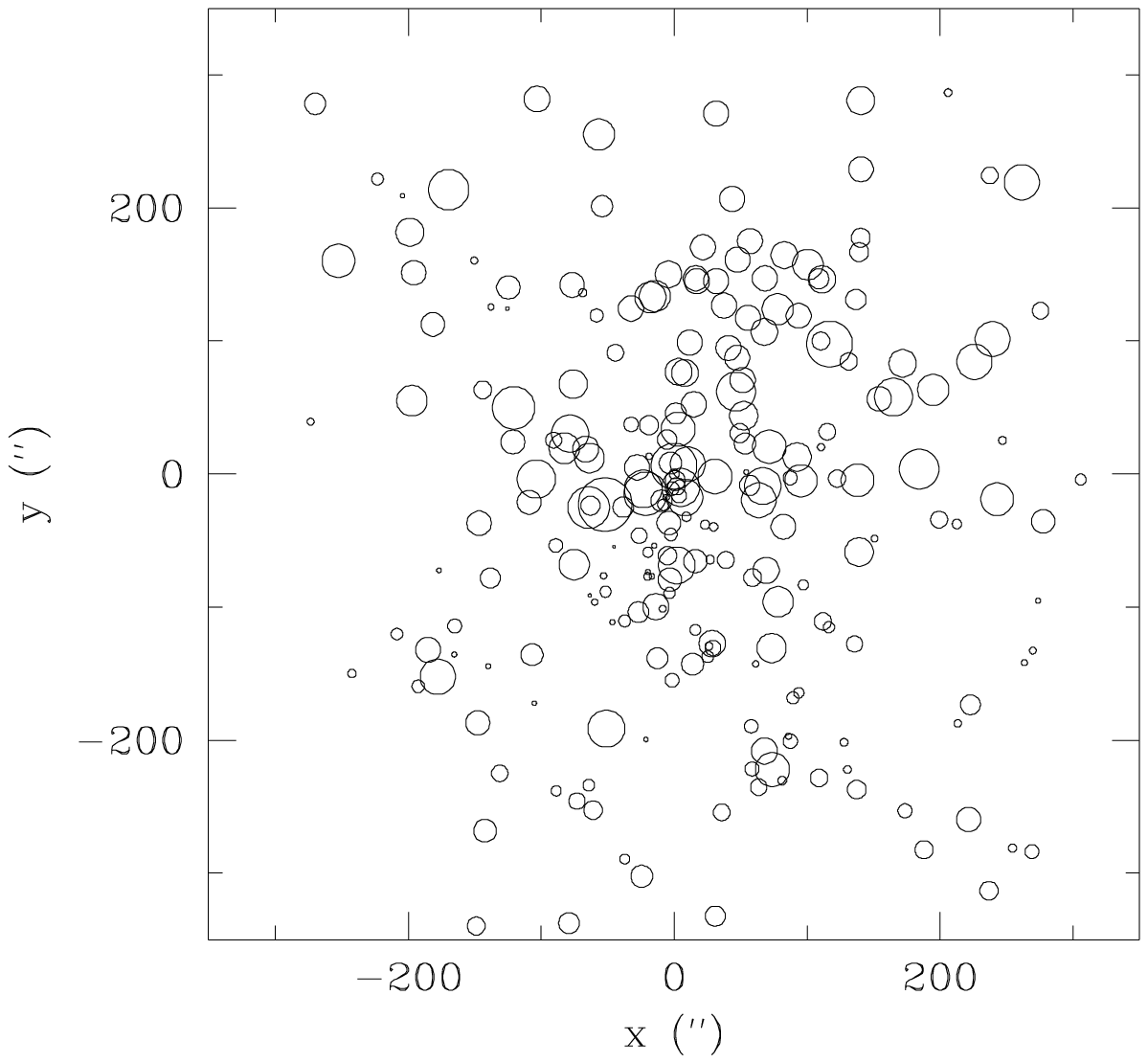]
{Positional map of the Dressler-Shectman statistic for all galaxies in our
catalog of CL 1358+62. The position of each galaxy is marked by a circle whose
diameter scales linearly with the deviation of the local kinematics, as
determined from the galaxy and its ten nearest neighbors, from the global
kinematics. The concentrations of large circles in the regions of the cluster
center and towards the upper-right quadrant indicate correlated spatial and
kinematic variations.}

\figcaption[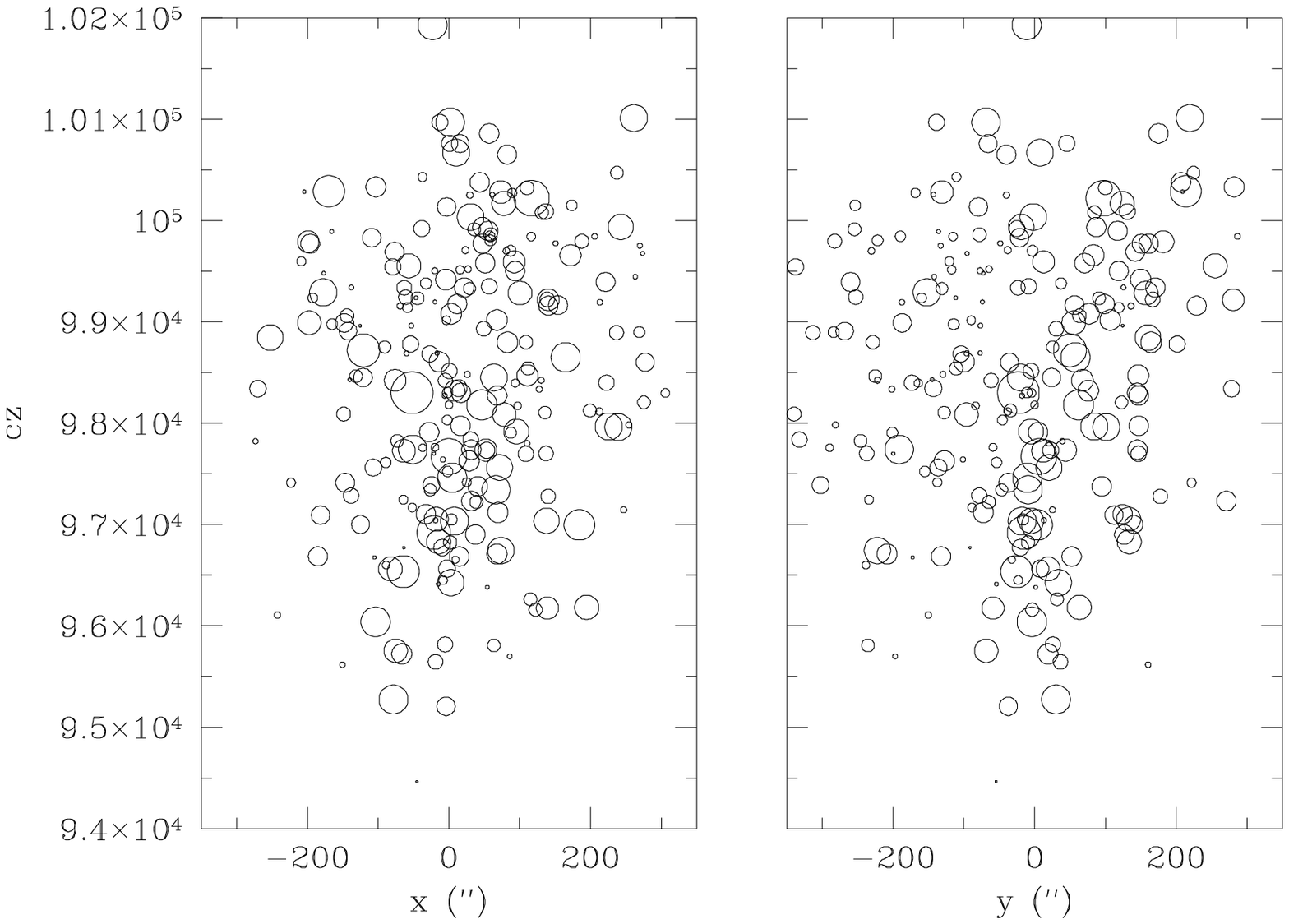]
{Position-velocity diagrams of the Dressler-Shectman statistic for CL 1358+62.
The narrow distribution of galaxies found near the cluster center ($x$ and $y$
$<$ 100\arcsec) and stretching from $cz$=95000 to $\sim$97000 km s$^{-1}$
constitutes a foreground group. At redshifts beyond 99000 km s$^{-1}$ a
concentration of galaxies is found at $y$$\approx$100\arcsec~and 
$x$$\approx$50\arcsec.}

\figcaption[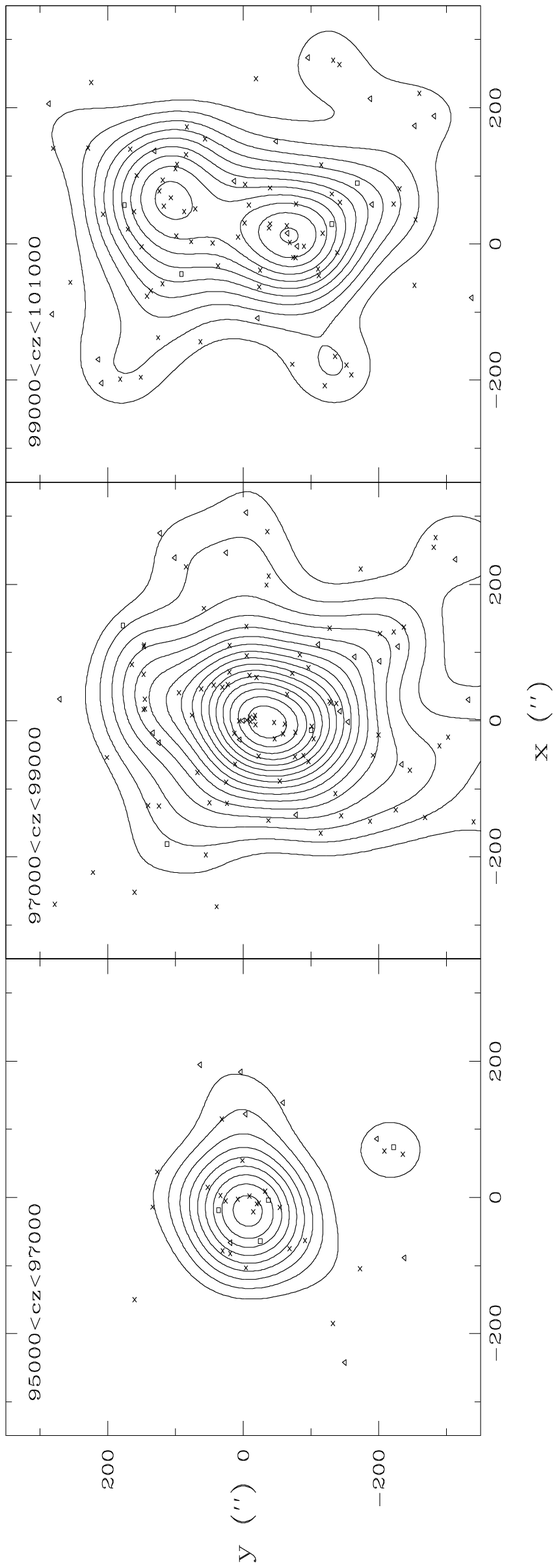]
{Redshift channels showing the distribution of cluster members in both spatial
and velocity dimensions. The channels are centered on the indicated redshifts
and spaced at intervals of 2000 km s$^{-1}$. Absorption-line galaxies are 
plotted as {\it crosses}, poststarburst galaxies are shown as {\it squares}, 
and emission-line galaxies are shown as {\it triangles}. Contours of galaxy
surface number density are shown. All panels have the same contour settings.
The cluster can be roughly divided into three main segments: a foreground
group at 95000$<$$cz$$<$97000 km s$^{-1}$, the main cluster component at 
97000$<$$cz$$<$99000 km s$^{-1}$, and an empty core with a background group
at 99000$<$$cz$$<$101000 km s$^{-1}$.}

\begin{deluxetable}{rrrccccrrrrc}
\scriptsize
\tablewidth{0pc}
\tablenum{1}
\tablecaption{CL 1358+62 Catalog}
\singlespace
\tablehead{
\colhead{Num} &
\colhead{x($^{\prime\prime}$)} &
\colhead{y($^{\prime\prime}$)} &
\colhead{$R$} &
\colhead{$V$$-$$R$} &
\colhead{$z$} &
\colhead{$\varepsilon$($z$)} &
\colhead{H$\delta$} &
\colhead{H$\gamma$} &
\colhead{H$\beta$} &
\colhead{[OII]} &
\colhead{Notes}
}
\startdata
    5 & $-$148.78 & $-$339.68 & 20.28 & \nodata &  0.32720 &  0.00020 &  $-$0.6$\pm$2.3 &  $-$0.2$\pm$4.0 &   1.2$\pm$1.8 &   1.1$\pm$3.5 & abs  \nl
    7 &  $-$78.99 & $-$337.52 & 20.71 & \nodata &  0.33204 &  0.00020 &   1.9$\pm$4.2 &   0.1$\pm$3.9 &  $-$0.1$\pm$3.1 & $-$17.3$\pm$4.8 & emi  \nl
   14 &   30.86 & $-$332.21 & 19.96 & \nodata &  0.32636 &  0.00020 &   3.7$\pm$2.1 &   3.1$\pm$2.8 &  $-$0.1$\pm$1.9 &  $-$7.1$\pm$2.2 & emi \nl
   28 &  236.73 & $-$313.15 & 21.36 & \nodata &  0.32987 &  0.00020 &   6.2$\pm$7.6 &  $-$0.3$\pm$7.2 &   0.7$\pm$4.2 & $-$25.7$\pm$9.7 & emi \nl
   43 &  $-$24.43 & $-$302.29 & 20.10 & \nodata &  0.32486 &  0.00020 &  $-$0.7$\pm$  2.1 &  $-$3.3$\pm$1.0 & 2.3$\pm$1.2 &  $-$0.3$\pm$2.3 &  abs  \nl
   49 &  $-$37.09 & $-$289.42 & 20.58 & \nodata &  0.32609 &  0.00020 &   1.2$\pm$ 4.6 &   1.2$\pm$6.1 &   3.5$\pm$2.5 &   1.4$\pm$4.0 &  abs  \nl
   53 &  268.95 & $-$283.77 & 20.42 & \nodata &  0.32989 &  0.00020 &  $-$2.3$\pm$ 1.6 &  $-$0.5$\pm$3.0 &   1.5$\pm$1.6 &   1.1$\pm$3.1 &  abs  \nl
   55 &  187.82 & $-$282.34 & 20.66 & \nodata &  0.33289 &  0.00038 & 6.4$\pm$3.2 & 0.6$\pm$4.0 & $-$0.9$\pm$2.1 & $-$16.1$\pm$5.0 & emi+H$\delta$ \nl
   57 &  254.48 & $-$281.28 & 20.66 & \nodata &  0.32684 &  0.00020 &  $-$6.9$\pm$ 7.2 &   0.1$\pm$5.7 &   2.3$\pm$2.4 &  0.4$\pm$18.4 &  abs  \nl
   77 & $-$142.29 & $-$268.16 & 21.56 & \nodata &  0.32993 &  0.00034 &   4.1$\pm$ 4.9 &   1.0$\pm$3.7 &  $-$0.2$\pm$4.3 &  $-$0.2$\pm$4.2 &  abs  \nl
   78 &  221.26 & $-$259.74 & 21.07 & \nodata &  0.33154 &  0.00020 &   1.2$\pm$3.5 &   3.0$\pm$5.3 &   5.6$\pm$1.5 &   5.4$\pm$7.0 &  abs  \nl
   86 &  $-$60.93 & $-$252.59 & 21.72 & \nodata &  0.33105 &  0.00023 &   1.0$\pm$ 3.1 &   2.5$\pm$2.2 &   3.0$\pm$2.1 &  $-$6.4$\pm$7.0 & abs  \nl
   89 &  $-$73.05 & $-$245.80 & 20.06 & \nodata &  0.32631 &  0.00020 &  $-$1.3$\pm$  1.8 &   0.7$\pm$2.0 &   2.6$\pm$1.4 &  $-$0.8$\pm$2.5 &  abs  \nl
   92 &  137.23 & $-$237.13 & 20.80 & \nodata &  0.32590 &  0.00025 &   5.7$\pm$3.0 &   2.6$\pm$3.2 &   2.3$\pm$2.6 &   0.0$\pm$2.5 &  abs  \nl
   95 &   63.53 & $-$235.36 & 20.48 &   0.89 &  0.31958 &  0.00020 &   1.1$\pm$3.4 &   5.7$\pm$3.2 &   2.0$\pm$2.2 &   0.6$\pm$4.4 &  abs  \nl
   97 &  $-$64.22 & $-$233.80 & 20.75 &         0.73 &  0.32437 &  0.00023 &   1.5$\pm$5.4 &   2.2$\pm$4.0 &   0.4$\pm$2.2 & $-$12.9$\pm$3.6 & emi  \nl
  100 &   81.23 & $-$230.40 & 19.79 &   0.85 &  0.33256 &  0.00020 &   0.9$\pm$2.0 &   1.5$\pm$3.0 &   3.5$\pm$1.6 &   0.2$\pm$3.1 &  abs  \nl
  105 &  130.37 & $-$222.09 & 21.46 &   0.65 &  0.32831 &  0.00020 &   0.8$\pm$4.7 &   1.3$\pm$2.4 &   3.1$\pm$1.5 &  $-$5.8$\pm$6.9 & abs \nl
  108 & $-$131.10 & $-$224.99 & 20.65 &         0.92 &  0.32844 &  0.00020 &  $-$0.7$\pm$2.2 &   0.4$\pm$2.8 &   2.8$\pm$1.3 &   1.7$\pm$2.6 &  abs  \nl
  109 &   73.68 & $-$222.07 & 21.43 &   0.67 &  0.32270 &  0.00020 &   8.1$\pm$2.8 &   7.5$\pm$2.9 &   5.3$\pm$2.0 &  $-$0.1$\pm$4.3 & e+a  \nl
  110 &   58.61 & $-$221.72 & 20.09 &   0.91 &  0.33291 &  0.00020 &  $-$0.5$\pm$2.0 &   0.9$\pm$1.8 &   3.0$\pm$1.5 &  $-$0.6$\pm$3.5 &  abs  \nl
  126 &   67.93 & $-$208.17 & 20.18 &   0.90 &  0.32259 &  0.00020 &  $-$1.0$\pm$1.3 &   0.3$\pm$2.4 &   2.8$\pm$1.6 &   2.1$\pm$1.8 &  abs  \nl
  129 &  127.68 & $-$201.66 & 20.10 &   0.82 &  0.32801 &  0.00020 &   0.0$\pm$1.7 &   1.0$\pm$1.4 &   2.6$\pm$0.7 &  $-$3.9$\pm$1.5 &  abs  \nl
  132 &   87.39 & $-$200.89 & 22.17 &   0.61 &  0.32658 &  0.00020 &   6.5$\pm$3.9 &   6.8$\pm$4.9 &  $-$2.3$\pm$3.0 & $-$12.8$\pm$5.5 & emi+H$\delta$  \nl
  135 &  $-$21.15 & $-$199.51 & 20.73 &         0.86 &  0.32590 &  0.00020 &  $-$0.6$\pm$3.1 & 1.4$\pm$3.5 & 4.0$\pm$2.8 &  $-$4.3$\pm$4.9 & abs  \nl
  137 &   86.00 & $-$196.95 & 21.58 &   0.76 &  0.31922 &  0.00029 &   8.0$\pm$5.5 &   0.1$\pm$9.9 &   4.1$\pm$3.2 & $-$12.8$\pm$7.6 & emi+H$\delta$ \nl
  142 &  $-$50.94 & $-$191.34 & 20.35 &         0.95 &  0.32603 &  0.00020 &   0.8$\pm$4.4 &  $-$0.2$\pm$3.2 &   2.6$\pm$1.6 &   1.0$\pm$2.8 &  abs  \nl
  143 &   57.97 & $-$189.63 & 21.05 &   0.87 &  0.33305 &  0.00020 &   0.7$\pm$3.1 &   4.3$\pm$4.5 &   3.3$\pm$2.9 &  $-$8.5$\pm$5.9 & emi \nl
  145 &  213.24 & $-$187.43 & 21.08 & \nodata &  0.33088 &  0.00034 &  $-$3.3$\pm$ 4.0 &  $-$8.9$\pm$6.3 & $-$13.2$\pm$ 4.0 & $-$53.9$\pm$5.6 & emi  \nl
  149 & $-$147.60 & $-$187.00 & 20.97 &         0.96 &  0.33020 &  0.00020 &  $-$1.2$\pm$3.7 &   0.4$\pm$4.5 &   2.1$\pm$1.5 &   4.0$\pm$3.1 & abs  \nl
  158 &  222.69 & $-$173.34 & 20.07 & \nodata &  0.32822 &  0.00020 &  $-$0.9$\pm$ 1.3 &  $-$0.5$\pm$1.7 &   3.1$\pm$1.1 &   0.3$\pm$1.9 &  abs  \nl
  164 & $-$105.06 & $-$172.23 & 19.95 &         0.95 &  0.32248 &  0.00020 &  $-$1.4$\pm$3.7 &   0.2$\pm$2.9 &   2.3$\pm$1.4 &   0.9$\pm$3.0 &  abs  \nl
  167 &   89.31 & $-$168.19 & 21.53 &   0.65 &  0.33447 &  0.00023 &   9.7$\pm$4.4 &  10.2$\pm$3.8 &   3.0$\pm$3.5 &  $-$3.6$\pm$6.1 & e+a \nl
  171 &   93.66 & $-$164.35 & 21.74 &   0.65 &  0.32822 &  0.00020 &   2.6$\pm$4.3 &   5.4$\pm$3.9 &  $-$1.4$\pm$2.7 &  $-$7.8$\pm$4.4 & emi \nl
  178 &   $-$1.61 & $-$155.08 & 20.57 &         0.51 &  0.32531 &  0.00020 &   5.4$\pm$3.6 & 4.3$\pm$2.2 & $-$9.0$\pm$3.5 & $-$34.2$\pm$1.7 & emi+H$\delta$  \nl
  182 & $-$177.65 & $-$152.25 & 21.07 &         0.91 &  0.33120 &  0.00023 &  $-$1.3$\pm$3.8 &   2.1$\pm$4.2 &   1.9$\pm$2.4 & $-$0.7$\pm$3.5 & abs  \nl
  187 &  263.32 & $-$141.80 & 21.75 & \nodata &  0.33172 &  0.00028 &   3.5$\pm$4.4 &   0.3$\pm$6.5 &   2.8$\pm$6.0 &   5.3$\pm$9.1 &  abs  \nl
  190 &   61.33 & $-$142.74 & 21.80 &   0.69 &  0.33442 &  0.00031 &   6.9$\pm$2.4 &   1.7$\pm$5.2 &   3.0$\pm$2.7 &  $-$3.6$\pm$3.8 &  abs  \nl
  192 &   13.83 & $-$142.88 & 21.27 &   0.63 &  0.32804 &  0.00020 &   2.9$\pm$3.6 &   1.5$\pm$6.8 &  $-$3.8$\pm$6.2 & $-$34.0$\pm$3.5 & emi \nl
  200 &  $-$12.63 & $-$138.44 & 20.04 &         0.81 &  0.33680 &  0.00040 &   4.8$\pm$4.0 &  $-$0.3$\pm$6.8 &   5.6$\pm$2.8 & $-$4.0$\pm$5.9 & abs  \nl
  203 &   25.25 & $-$137.36 & 20.86 &   0.91 &  0.32495 &  0.00020 &   1.5$\pm$4.6 &   3.8$\pm$3.7 &   3.6$\pm$2.7 &   2.2$\pm$4.0 &  abs  \nl
  204 &  269.67 & $-$132.66 & 19.74 & \nodata &  0.33274 &  0.00020 &  $-$0.1$\pm$ 2.3 &  $-$1.2$\pm$2.9 &   1.9$\pm$1.4 &   3.7$\pm$2.0 &  abs  \nl
  205 & $-$106.81 & $-$135.88 & 21.76 &         0.89 &  0.32544 &  0.00020 &   5.5$\pm$6.1 &   2.0$\pm$5.7 &   1.7$\pm$2.2 &   7.3$\pm$5.1 &  abs  \nl
  206 & $-$165.28 & $-$135.64 & 21.48 &         0.82 &  0.33321 &  0.00027 &   7.2$\pm$2.0 &   1.6$\pm$4.9 &   2.5$\pm$2.1 &  $-$7.5$\pm$7.6 & abs  \nl
  207 &   73.35 & $-$130.84 & 21.86 &   0.78 &  0.33450 &  0.00020 &   0.9$\pm$3.6 &  $-$0.8$\pm$7.5 &   2.7$\pm$2.1 &  $-$0.1$\pm$6.1 &  abs  \nl
  209 &   29.25 & $-$130.76 & 19.97 &   0.72 &  0.33133 &  0.00020 &   5.9$\pm$1.1 &   6.9$\pm$1.6 &   5.2$\pm$1.3 &  $-$0.8$\pm$2.8 & e+a \nl
  210 & $-$185.02 & $-$132.14 & 20.39 &         1.02 &  0.32251 &  0.00020 &  $-$2.7$\pm$2.9 &   1.3$\pm$2.1 &   2.6$\pm$1.6 &   3.2$\pm$4.2 &  abs  \nl
  211 &   26.11 & $-$129.35 & 20.10 &   0.91 &  0.32850 &  0.00020 &   0.7$\pm$1.3 &   2.2$\pm$3.0 &   2.4$\pm$0.7 &   0.7$\pm$1.4 &  abs  \nl
  212 &  135.61 & $-$127.67 & 20.16 &   0.93 &  0.32724 &  0.00020 &  $-$2.2$\pm$3.5 &  $-$0.8$\pm$3.9 &   2.7$\pm$1.5 &  $-$2.9$\pm$6.5 &  abs  \nl
  215 &   28.54 & $-$127.40 & 20.71 &   1.01 &  0.32565 &  0.00020 &  $-$0.3$\pm$4.5 &   2.7$\pm$3.6 &   1.9$\pm$2.5 &   0.7$\pm$3.7 &  abs  \nl
  226 & $-$208.35 & $-$120.30 & 21.26 &         0.70 &  0.33222 &  0.00020 &   4.8$\pm$2.7 &   2.9$\pm$4.0 &   3.3$\pm$1.8 &   0.8$\pm$3.5 &  abs  \nl
  227 &   15.91 & $-$117.08 & 21.00 &   0.96 &  0.33194 &  0.00022 &   1.2$\pm$2.7 &   2.9$\pm$4.8 &   2.4$\pm$1.4 &   0.5$\pm$5.6 &  abs  \nl
  233 & $-$164.92 & $-$114.30 & 19.49 &         0.97 &  0.33015 &  0.00020 &  $-$0.9$\pm$1.6 &   0.4$\pm$1.4 &   3.3$\pm$0.9 &   1.4$\pm$1.9 &  abs  \nl
  234 &  111.84 & $-$110.76 & 20.13 &   0.73 &  0.32870 &  0.00020 &   1.9$\pm$1.8 &   1.3$\pm$3.0 &  $-$3.1$\pm$1.8 & $-$14.3$\pm$2.4 & emi \nl
  235 &  $-$46.46 & $-$111.33 & 21.55 &         0.87 &  0.33102 &  0.00022 &   1.0$\pm$3.3 &   3.1$\pm$3.8 &   0.2$\pm$2.8 &   1.3$\pm$6.5 &  abs  \nl
  236 &  $-$37.20 & $-$110.38 & 20.22 &         0.99 &  0.33500 &  0.00027 &   1.0$\pm$1.8 &   0.1$\pm$5.7 &   3.1$\pm$2.0 &   4.6$\pm$3.2 &  abs  \nl
  239 &  $-$26.80 & $-$103.88 & 20.72 &         0.99 &  0.32917 &  0.00020 &  $-$1.0$\pm$3.2 &   1.2$\pm$3.0 &   3.3$\pm$2.2 &   3.7$\pm$4.1 &  abs  \nl
  242 &   $-$8.55 & $-$101.16 & 19.88 &         0.95 &  0.32570 &  0.00027 &   0.3$\pm$3.3 &   0.0$\pm$3.8 &   2.0$\pm$1.8 &   5.0$\pm$2.6 &  abs  \nl
  243 &  $-$13.78 &  $-$99.71 & 21.29 &         0.62 &  0.32892 &  0.00020 &  10.4$\pm$1.3 &  10.9$\pm$1.7 &   6.0$\pm$1.7 &   2.1$\pm$3.4 & e+a \nl
  244 &  273.59 &  $-$95.30 & 20.87 & \nodata &  0.33248 &  0.00020 &   2.3$\pm$1.5 &   2.4$\pm$2.3 &  $-$0.2$\pm$1.5 & $-$11.4$\pm$1.2 & emi \nl
  246 &   78.11 &  $-$96.07 & 21.08 &   0.92 &  0.32718 &  0.00020 &   3.4$\pm$ 4.0 &   3.5$\pm$3.4 &   3.1$\pm$2.2 &   3.4$\pm$8.4 &  abs  \nl
  248 &  $-$59.75 &  $-$96.18 & 20.32 & 1.05 &  0.32919 &  0.00020 &  $-$1.6$\pm$2.5 &  $-$1.1$\pm$2.4 &   2.7$\pm$0.9 &   2.5$\pm$4.7 &  abs  \nl
  254 &  $-$63.46 &  $-$91.22 & 20.01 & 0.97 &  0.32280 &  0.00027 &  $-$0.4$\pm$3.0 &  $-$1.6$\pm$3.1 &   3.1$\pm$1.5 &   1.9$\pm$3.6 &  abs  \nl
  256 &  $-$51.62 &  $-$88.49 & 19.00 & 1.00 &  0.32412 &  0.00041 &   3.0$\pm$ 1.1 &   0.1$\pm$2.1 &   2.5$\pm$1.3 &   0.8$\pm$2.7 &  abs  \nl
  264 &   $-$3.13 &  $-$79.26 & 22.12 & 0.42 &  0.33401 &  0.00020 &   3.3$\pm$ 2.5 &  $-$1.8$\pm$3.2 & $-$14.3$\pm$3.4 & $-$58.8$\pm$5.6 & emi \nl
  265 &   58.95 &  $-$77.87 & 20.57 &   0.96 &  0.33310 &  0.00040 &   5.0$\pm$ 3.0 &   1.3$\pm$4.6 &   1.0$\pm$1.8 &   0.7$\pm$1.6 &  abs  \nl
  269 &  $-$16.73 &  $-$76.88 & 19.12 & 1.02 &  0.32920 &  0.00027 &   0.1$\pm$ 1.6 &   0.4$\pm$2.9 &   2.6$\pm$1.4 &   0.7$\pm$1.7 &  abs  \nl
  271 & $-$138.19 &  $-$78.11 & 21.78 & 0.54 &  0.32451 &  0.00027 &   4.5$\pm$ 3.3 &  $-$1.2$\pm$2.2 & $-$13.0$\pm$4.1 & $-$55.1$\pm$4.8 & emi+H$\delta$ \nl
  272 &  $-$53.03 &  $-$76.45 & 20.99 & 0.93 &  0.33010 &  0.00040 &  $-$3.6$\pm$4.3 &   1.8$\pm$4.3 &   0.5$\pm$1.3 &  $-$3.5$\pm$6.0 &  abs  \nl
  278 &  $-$19.71 &  $-$73.64 & 20.57 & 0.99 &  0.33089 &  0.00044 &   0.1$\pm$ 2.5 &  $-$1.4$\pm$2.3 &   2.4$\pm$0.7 &   2.3$\pm$3.4 &  abs  \nl
  279 &   69.25 &  $-$72.37 & 20.57 &   0.97 &  0.32396 &  0.00055 &   1.6$\pm$ 3.7 &   1.1$\pm$4.2 &   4.8$\pm$2.5 &   1.8$\pm$3.1 &  abs  \nl
  283 & $-$176.74 &  $-$72.38 & 20.54 & 0.92 &  0.33183 &  0.00020 &  $-$0.1$\pm$2.8 &   1.3$\pm$3.8 &   3.2$\pm$1.3 &  $-$0.3$\pm$3.8 &  abs  \nl
  285 &    1.91 &  $-$68.62 & 21.58 &   1.01 &  0.33680 &  0.00040 &  $-$9.3$\pm$8.9 &  $-$2.7$\pm$7.8 &  $-$0.6$\pm$5.2 &  7.1$\pm$13.0 &  abs  \nl
  288 &  $-$75.07 &  $-$68.20 & 20.69 & 0.99 &  0.31940 &  0.00027 &  $-$1.3$\pm$4.0 &   1.4$\pm$3.8 &   0.1$\pm$1.3 &   0.2$\pm$8.9 &  abs  \nl
  290 &   15.96 &  $-$65.50 & 21.08 &   0.74 &  0.33610 &  0.00040 &   3.7$\pm$ 1.1 &   1.7$\pm$5.4 &   3.6$\pm$1.4 &  $-$6.9$\pm$3.4 &  emi  \nl
  292 &   38.86 &  $-$64.58 & 20.09 &   0.92 &  0.32430 &  0.00027 &   1.9$\pm$ 2.6 &   1.6$\pm$2.4 &   4.2$\pm$1.1 &   1.5$\pm$4.0 &  abs  \nl
  293 &   27.06 &  $-$64.30 & 21.61 &   1.01 &  0.33197 &  0.00020 &   1.3$\pm$ 3.3 &   1.4$\pm$2.0 &   2.2$\pm$2.1 &   0.7$\pm$4.0 &  abs  \nl
  295 &   $-$4.97 &  $-$61.35 & 20.18 & 0.96 &  0.32830 &  0.00027 &   0.1$\pm$ 4.2 &   0.8$\pm$5.6 &   4.2$\pm$2.9 &   1.8$\pm$3.6 &  abs  \nl
  298 &  $-$19.71 &  $-$58.77 & 19.46 & 1.03 &  0.32610 &  0.00027 &  $-$2.8$\pm$2.3 &   0.8$\pm$2.3 &   2.8$\pm$0.8 &  $-$0.7$\pm$6.0 &  abs  \nl
  299 &  $-$45.12 &  $-$54.61 & 21.37 & 0.89 &  0.31510 &  0.00021 &   0.2$\pm$ 4.0 &   1.5$\pm$2.4 &   1.2$\pm$0.9 &   4.4$\pm$3.6 &  abs  \nl
  300 &  $-$14.97 &  $-$53.87 & 20.03 & 0.98 &  0.32160 &  0.00027 &   0.8$\pm$ 2.1 &   2.6$\pm$2.8 &   1.8$\pm$1.8 &  $-$4.7$\pm$3.8 &  abs  \nl
  303 &  $-$88.96 &  $-$53.79 & 20.04 & 0.98 &  0.32560 &  0.00027 &   0.9$\pm$ 1.7 &  $-$1.6$\pm$2.9 &   2.8$\pm$1.5 &   0.0$\pm$4.3 &  abs  \nl
  305 &  150.68 &  $-$48.39 & 21.00 &   0.86 &  0.33282 &  0.00020 &   3.6$\pm$ 3.4 &   2.5$\pm$3.2 &   0.8$\pm$2.1 &  $-$5.5$\pm$5.3 & emi \nl
  309 &  $-$26.33 &  $-$46.29 & 19.86 & 1.03 &  0.32470 &  0.00027 &  $-$1.2$\pm$2.2 &   2.2$\pm$4.1 &   1.9$\pm$1.1 & \nodata &  abs  \nl
  311 &   $-$2.50 &  $-$45.45 & 21.20 & 0.98 &  0.32700 &  0.00040 &   1.1$\pm$ 7.0 &   7.3$\pm$7.2 &   4.2$\pm$3.1 &  $-$5.8$\pm$9.9 & abs \nl
  318 &  212.54 &  $-$37.61 & 21.00 &   0.89 &  0.32728 &  0.00020 &   1.0$\pm$ 3.6 &   3.5$\pm$4.1 &   3.9$\pm$1.4 &  $-$1.9$\pm$4.6 &  abs  \nl
  319 &   29.68 &  $-$39.69 & 20.72 &   0.97 &  0.33440 &  0.00040 &  $-$0.6$\pm$4.6 &   2.5$\pm$6.7 &   3.5$\pm$3.2 &   2.1$\pm$5.0 &  abs  \nl
  321 &  277.41 &  $-$35.51 & 20.08 & \nodata &  0.32890 &  0.00020 &   3.7$\pm$1.8 &   2.6$\pm$3.2 &   3.2$\pm$1.6 &   2.9$\pm$2.0 &  abs  \nl
  323 &   23.29 &  $-$38.10 & 21.11 &   0.90 &  0.33259 &  0.00020 &   0.5$\pm$ 5.0 &   2.5$\pm$3.7 &   2.2$\pm$1.8 & $-$1.6$\pm$13.9 &  abs  \nl
  328 &   $-$4.03 &  $-$36.82 & 19.96 & 0.72 &  0.31757 &  0.00039 &   7.2$\pm$ 1.3 &   6.6$\pm$2.2 &   3.8$\pm$2.1 &  $-$1.0$\pm$2.4 & e+a \nl
  329 &  199.19 &  $-$34.23 & 20.72 &   0.93 &  0.32731 &  0.00020 &   0.8$\pm$ 1.6 &   1.5$\pm$1.5 &   4.2$\pm$1.3 &  $-$0.6$\pm$4.4 &  abs  \nl
  331 & $-$146.58 &  $-$36.92 & 21.95 & 0.90 &  0.32493 &  0.00020 &  $-$5.0$\pm$5.3 &   1.2$\pm$4.2 &   1.2$\pm$2.1 & $-$3.6$\pm$11.0 &  abs  \nl
  335 &    9.21 &  $-$32.17 & 20.57 &   0.95 &  0.32240 &  0.00027 &   1.8$\pm$ 3.9 &   3.2$\pm$3.5 &   2.1$\pm$2.4 &$-$14.8$\pm$12.0 & abs \nl
  343 &  $-$64.33 &  $-$25.27 & 20.64 & 0.78 &  0.32201 &  0.00027 &   5.3$\pm$ 2.9 &   8.4$\pm$3.9 &   4.8$\pm$2.4 &   1.0$\pm$5.0 & e+a \nl
  344 &  $-$38.25 &  $-$24.87 & 20.85 & 0.96 &  0.33330 &  0.00027 &   2.3$\pm$ 4.3 &   2.7$\pm$8.2 &   5.0$\pm$2.9 &  $-$1.3$\pm$4.6 &  abs  \nl
  346 &   $-$8.17 &  $-$23.07 & 19.78 & 0.89 &  0.32173 &  0.00020 &   2.2$\pm$ 1.9 &   3.1$\pm$2.0 &   3.7$\pm$0.9 &  $-$0.8$\pm$2.4 &  abs  \nl
  347 &  $-$51.92 &  $-$22.89 & 20.13 & 0.97 &  0.32790 &  0.00040 &   4.2$\pm$ 3.1 &  $-$0.7$\pm$3.4 &  $-$0.2$\pm$2.7 &   1.0$\pm$5.6 &  abs  \nl
  348 &  242.81 &  $-$18.89 & 20.50 &   0.94 &  0.33336 &  0.00020 &  $-$0.3$\pm$3.9 &  $-$2.6$\pm$3.1 &   2.8$\pm$2.5 &   0.2$\pm$5.1 &  abs  \nl
  349 &   63.65 &  $-$19.66 & 21.47 &   0.93 &  0.32840 &  0.00040 &   4.4$\pm$ 6.9 &  $-$0.9$\pm$11.2 &   2.6$\pm$ 3.2 &   1.2$\pm$5.6 &  abs  \nl
  350 &   $-$9.53 &  $-$20.00 & 21.06 & 0.97 &  0.32280 &  0.00040 &   3.5$\pm$ 4.5 &   0.4$\pm$12.9 &   2.1$\pm$ 1.7 &   6.4$\pm$3.0 &  abs  \nl
  351 & $-$109.02 &  $-$21.23 & 20.10 & 0.72 &  0.33300 &  0.00040 &   3.5$\pm$ 2.4 &   1.4$\pm$3.7 &  $-$0.3$\pm$2.0 &  $-$6.5$\pm$2.1 & emi \nl
  352 &   $-$6.09 &  $-$17.74 & 20.95 & 1.01 &  0.32780 &  0.00040 &  $-$0.9$\pm$8.5 &   2.9$\pm$9.4 &  $-$1.7$\pm$4.2 &   9.5$\pm$4.8 &  abs  \nl
  353 &    7.80 &  $-$17.57 & 19.42 &   1.03 &  0.32367 &  0.00041 &  $-$0.9$\pm$1.3 &   2.1$\pm$2.0 &   1.8$\pm$0.8 &   0.7$\pm$1.8 &  abs  \nl
  354 &    3.73 &  $-$16.28 & 20.98 &   1.00 &  0.32372 &  0.00028 &   4.6$\pm$ 4.7 &   3.3$\pm$4.1 &   1.8$\pm$2.2 &  $-$4.9$\pm$6.3 &  abs  \nl
  356 &  $-$23.23 &  $-$11.19 & 19.68 & 1.06 &  0.34000 &  0.00040 &   2.7$\pm$ 6.2 &  $-$5.3$\pm$7.2 &   2.6$\pm$3.7 &$-$20.2$\pm$15.2 & abs \nl
  357 &   $-$1.58 &  $-$10.79 & 20.83 & 0.98 &  0.32790 &  0.00040 &   5.4$\pm$ 8.4 &   0.8$\pm$15.5 &  $-$3.8$\pm$ 4.9 & $-$11.2$\pm$9.1 & abs \nl
  358 &    4.91 &  $-$10.15 & 20.83 &   0.99 &  0.32510 &  0.00040 &  $-$2.0$\pm$4.7 &  $-$4.2$\pm$6.9 &   4.9$\pm$2.4 &   8.8$\pm$5.3 &  abs  \nl
  359 &   66.68 &   $-$9.02 & 19.91 &   0.97 &  0.32470 &  0.00027 &  $-$0.1$\pm$1.2 &   2.7$\pm$1.6 &   3.1$\pm$0.9 &   2.5$\pm$3.0 &  abs  \nl
  360 &    1.76 &   $-$9.25 & 20.62 &   0.98 &  0.32297 &  0.00020 &   1.3$\pm$ 2.4 &   1.1$\pm$2.6 &   3.1$\pm$1.4 &  $-$0.1$\pm$3.0 &  abs  \nl
  361 &   56.87 &   $-$8.42 & 21.03 &   0.94 &  0.33140 &  0.00040 &  $-$5.6$\pm$3.6 &   3.7$\pm$7.4 &   1.6$\pm$2.9 & \nodata &  abs  \nl
  362 &  305.72 &   $-$4.20 & 20.66 & \nodata &  0.32789 &  0.00024 &   4.7$\pm$1.8 &   3.9$\pm$1.6 &  $-$2.0$\pm$1.5 & $-$24.7$\pm$2.0 & emi+H$\delta$ \nl
  366 &  137.90 &   $-$4.65 & 20.25 &   1.00 &  0.32368 &  0.00020 &   0.8$\pm$ 2.6 &   3.2$\pm$2.5 &   3.3$\pm$1.5 &  $-$0.4$\pm$4.0 &  abs  \nl
  368 &    0.78 &   $-$5.02 & 20.31 &   1.01 &  0.32861 &  0.00059 &  $-$0.5$\pm$2.9 &   1.1$\pm$2.9 &   4.1$\pm$1.4 &  $-$0.2$\pm$3.9 &  abs  \nl
  369 &  122.54 &   $-$3.36 & 20.42 &   0.96 &  0.32076 &  0.00020 &   1.4$\pm$ 2.9 &   1.6$\pm$3.1 &   2.7$\pm$1.6 &  $-$6.1$\pm$3.9 & emi \nl
  371 &   87.25 &   $-$2.90 & 19.71 &   1.02 &  0.33257 &  0.00039 &  $-$1.3$\pm$1.5 &  $-$0.7$\pm$2.5 &   3.0$\pm$1.3 &  $-$2.1$\pm$2.0 &  abs  \nl
  372 & $-$103.57 &   $-$3.81 & 20.07 & 0.91 &  0.32035 &  0.00051 &   0.5$\pm$ 3.0 &   0.9$\pm$2.3 &   3.4$\pm$1.4 &   1.4$\pm$2.0 &  abs  \nl
  375 &    0.00 &    0.00 & 19.09 & 1.04 &  0.32750 &  0.00040 &  $-$2.9$\pm$3.9 &   1.4$\pm$3.9 &   2.6$\pm$2.3 & $-$11.3$\pm$4.1 & bcg+emi \nl
  376 &   54.14 &    1.47 & 20.48 & 1.00 &  0.32150 &  0.00027 &   0.1$\pm$ 2.7 &   2.1$\pm$5.0 &   2.5$\pm$1.3 & \nodata  &  abs  \nl
  377 &  184.21 &    3.80 & 20.40 & 0.69 &  0.32355 &  0.00020 &   3.5$\pm$ 3.2 &   4.0$\pm$4.1 &   0.8$\pm$2.5 &  $-$6.7$\pm$3.5 & emi \nl
  380 &  $-$27.82 &    4.98 & 20.80 &   0.54 &  0.32660 &  0.00040 &   5.1$\pm$2.6 & 5.8$\pm$5.6 & 0.0$\pm$3.4 & $-$11.3$\pm$5.6 & emi+H$\delta$ \nl
  381 &   $-$0.06 &    6.01 & 19.91 &   0.98 &  0.32580 &  0.00027 &   0.8$\pm$2.7 &   2.9$\pm$4.5 &   2.6$\pm$1.7 &   2.7$\pm$3.9 &  abs  \nl
  383 &   10.29 &    7.60 & 21.23 &     0.97 &  0.33580 &  0.00104 &  $-$0.3$\pm$2.9 &   3.1$\pm$5.9 &   0.6$\pm$2.4 &   0.3$\pm$2.7 &  abs  \nl
  386 &   $-$2.73 &    8.27 & 20.68 &   1.11 &  0.32210 &  0.00020 &  $-$6.4$\pm$5.9 &  $-$0.4$\pm$6.5 &   3.9$\pm$3.7 &   6.3$\pm$5.3 &  abs  \nl
  388 &  $-$63.65 &   12.02 & 21.95 &   0.93 &  0.32599 &  0.00020 &  $-$0.4$\pm$5.2 &  $-$1.7$\pm$4.3 &   2.0$\pm$2.9 &   1.2$\pm$4.7 &  abs  \nl
  391 &  $-$18.77 &   13.22 & 19.81 &   0.99 &  0.32370 &  0.00027 &  $-$0.1$\pm$4.2 &   0.2$\pm$4.1 &   0.6$\pm$1.9 &   0.1$\pm$4.7 &  abs  \nl
  394 &  110.45 &   20.22 & 21.34 &     0.96 &  0.32623 &  0.00028 &   0.6$\pm$4.2 &   2.4$\pm$4.4 &   1.0$\pm$1.9 &  2.4$\pm$12.7 &  abs  \nl
  396 &  $-$66.44 &   18.82 & 20.95 &   0.51 &  0.31930 &  0.00027 &   4.0$\pm$2.0 &  $-$3.4$\pm$4.4 & $-$14.0$\pm$3.0 & $-$47.7$\pm$6.9 & emi+H$\delta$ \nl
  397 &  $-$82.60 &   19.35 & 20.61 &   0.96 &  0.32210 &  0.00027 &  $-$2.8$\pm$3.0 &   1.4$\pm$3.0 &   3.8$\pm$1.6 &   4.9$\pm$4.5 &  abs  \nl
  399 &  246.83 &   25.25 & 20.49 &     0.64 &  0.32404 &  0.00023 &   6.6$\pm$2.0 &   3.3$\pm$1.6 &  $-$1.2$\pm$1.6 & $-$11.9$\pm$2.1 & emi+H$\delta$ \nl
  400 &   53.22 &   22.94 & 21.54 &     0.94 &  0.32600 &  0.00040 &   7.2$\pm$9.7 &   0.4$\pm$12.9 &   2.5$\pm$3.2 & 16.5$\pm$11.8 &  abs  \nl
  406 & $-$121.17 &   24.15 & 21.82 &   0.88 &  0.32840 &  0.00030 &   3.0$\pm$6.5 &   4.0$\pm$2.8 &   2.4$\pm$4.2 &  $-$1.9$\pm$6.3 &  abs  \nl
  408 &   $-$5.33 &   25.92 & 20.25 &   0.99 &  0.31961 &  0.00041 &  $-$2.0$\pm$2.6 &   0.8$\pm$3.3 &   3.0$\pm$1.3 &   0.2$\pm$2.5 &  abs  \nl
  409 &  $-$90.55 &   25.48 & 21.16 &   0.99 &  0.32940 &  0.00040 &  $-$2.7$\pm$5.6 &   2.6$\pm$8.3 &   3.5$\pm$3.4 &   4.2$\pm$4.9 &  abs  \nl
  410 &   49.29 &   30.66 & 20.70 &     1.03 &  0.33000 &  0.00027 &  $-$3.4$\pm$2.7 &   1.2$\pm$2.9 &   2.1$\pm$1.2 &   0.8$\pm$2.2 &  abs  \nl
  411 &  115.12 &   31.73 & 21.04 &     1.01 &  0.32110 &  0.00040 &   4.0$\pm$3.1 &   0.3$\pm$5.3 &   3.3$\pm$2.2 &   3.8$\pm$5.8 &  abs  \nl
  412 &  $-$78.28 &   30.39 & 20.44 &   0.91 &  0.31780 &  0.00040 &  $-$0.1$\pm$4.2 &   2.3$\pm$5.3 &   6.2$\pm$3.1 &  $-$0.7$\pm$3.0 &  abs  \nl
  416 &    2.95 &   33.74 & 21.52 &     1.06 &  0.32165 &  0.00020 &  $-$0.1$\pm$4.0 &   1.9$\pm$3.8 &   1.2$\pm$2.5 &   4.5$\pm$6.8 &  abs  \nl
  420 &  $-$18.87 &   36.69 & 21.97 &   0.74 &  0.31904 &  0.00020 &   3.2$\pm$3.4 &   5.9$\pm$3.8 &   5.7$\pm$3.0 &   1.0$\pm$6.4 & e+a \nl
  421 &  $-$32.31 &   37.35 & 20.83 &   1.04 &  0.33150 &  0.00040 &  $-$1.7$\pm$2.4 &   0.5$\pm$5.1 &   2.9$\pm$1.4 &  $-$1.6$\pm$4.7 &  abs  \nl
  430 & $-$273.20 &   39.29 & 20.72 &   0.90 &  0.32630 &  0.00022 &   2.3$\pm$5.2 &   2.1$\pm$4.3 &   5.9$\pm$4.2 &  $-$0.6$\pm$2.2 &  abs  \nl
  433 &   52.28 &   43.85 & 20.16 &     0.96 &  0.32601 &  0.00059 &  $-$1.3$\pm$2.7 &   0.1$\pm$2.8 &   3.6$\pm$1.3 &   1.5$\pm$1.8 &  abs  \nl
  434 &    1.34 &   45.48 & 21.59 &     1.01 &  0.33611 &  0.00020 &  $-$1.8$\pm$1.9 &   0.5$\pm$2.9 &   2.5$\pm$1.8 &  $-$2.1$\pm$4.0 &  abs  \nl
  440 & $-$120.58 &   49.83 & 21.05 &   0.97 &  0.32929 &  0.00020 &   0.8$\pm$4.7 &   0.7$\pm$2.1 &   3.1$\pm$1.3 &  $-$0.1$\pm$4.8 &  abs  \nl
  442 &   14.85 &   52.33 & 21.20 &     0.92 &  0.32251 &  0.00020 &   0.9$\pm$4.6 &   4.1$\pm$5.1 &   0.9$\pm$1.8 &   7.7$\pm$7.8 &  abs  \nl
  444 &  154.20 &   56.42 & 20.29 &     1.04 &  0.33078 &  0.00020 &   0.4$\pm$2.8 &   1.8$\pm$2.6 &   1.9$\pm$1.2 &  $-$1.5$\pm$4.9 &  abs  \nl
  447 &  164.96 &   57.93 & 20.73 &     0.99 &  0.32906 &  0.00020 &   1.5$\pm$2.3 &   3.0$\pm$3.3 &   1.6$\pm$1.8 &   1.6$\pm$3.5 &  abs  \nl
  452 & $-$197.18 &   55.21 & 20.39 &   0.97 &  0.33020 &  0.00020 &   0.4$\pm$2.5 &   0.4$\pm$2.8 &   3.2$\pm$1.4 &  $-$1.8$\pm$4.5 &  abs  \nl
  453 &  194.73 &   63.33 & 20.40 &     0.56 &  0.32083 &  0.00020 &   6.1$\pm$2.5 &   3.8$\pm$1.9 &  $-$8.6$\pm$2.2 & $-$40.1$\pm$4.6 & emi+H$\delta$ \nl
  454 &   46.57 &   61.96 & 19.89 &     0.96 &  0.32750 &  0.00027 &   1.8$\pm$2.3 &   2.0$\pm$4.0 &   3.7$\pm$1.5 &  $-$0.8$\pm$1.7 &  abs  \nl
  457 & $-$143.89 &   63.30 & 21.02 &   0.99 &  0.33043 &  0.00020 &   3.9$\pm$3.7 &  $-$0.5$\pm$2.5 &   2.6$\pm$0.8 &  $-$3.3$\pm$7.1 &  abs  \nl
  460 &   51.45 &   70.69 & 21.36 &     1.00 &  0.33216 &  0.00020 &   0.5$\pm$2.9 &  $-$0.4$\pm$3.9 &   2.8$\pm$1.6 &  $-$1.2$\pm$4.4 &  abs  \nl
  463 &    8.03 &   75.87 & 19.91 &     0.99 &  0.32797 &  0.00059 &   1.1$\pm$1.6 &   1.7$\pm$2.9 &   2.3$\pm$1.5 &   3.7$\pm$2.4 &  abs  \nl
  465 &    3.38 &   76.79 & 20.53 &     1.00 &  0.33050 &  0.00040 &   3.7$\pm$2.7 &  $-$0.7$\pm$8.2 &   4.5$\pm$2.3 &$-$13.8$\pm$13.5 & abs \nl
  468 &  171.80 &   83.16 & 20.52 &     1.02 &  0.33242 &  0.00020 &   1.8$\pm$1.9 &  $-$1.5$\pm$3.1 &   2.3$\pm$1.6 &  $-$0.9$\pm$3.1 &  abs  \nl
  469 &  225.73 &   84.35 & 21.51 &     0.85 &  0.32679 &  0.00020 &   3.4$\pm$4.1 &  $-$0.9$\pm$2.9 &   1.8$\pm$2.2 &  $-$0.6$\pm$4.2 &  abs  \nl
  470 &  131.21 &   84.61 & 19.50 &     1.01 &  0.33383 &  0.00020 &   0.0$\pm$2.8 &  $-$0.2$\pm$1.8 &   3.2$\pm$0.9 &  $-$0.8$\pm$1.9 &  abs  \nl
  473 &   47.42 &   87.30 & 21.04 &     0.89 &  0.33335 &  0.00020 &   2.8$\pm$3.0 &   0.8$\pm$5.1 &   3.1$\pm$2.0 &   1.2$\pm$5.5 &  abs  \nl
  481 &  $-$44.17 &   91.14 & 21.24 &   0.95 &  0.33101 &  0.00064 &   4.1$\pm$4.1 &   4.2$\pm$3.9 &   4.4$\pm$2.4 &   6.9$\pm$3.6 & e+a \nl
  482 &   40.83 &   94.80 & 21.28 &     1.05 &  0.32481 &  0.00020 &   1.6$\pm$3.3 &   3.2$\pm$3.6 &   2.7$\pm$1.1 &   0.3$\pm$2.5 &  abs  \nl
  491 &  239.36 &  101.40 & 20.50 &     0.88 &  0.32677 &  0.00020 &   2.1$\pm$1.9 &   0.4$\pm$1.3 &  $-$0.9$\pm$1.0 & $-$12.4$\pm$1.4 & emi \nl
  492 &  110.31 &   99.96 & 20.61 &     1.04 &  0.33464 &  0.00020 &   2.4$\pm$1.7 &   3.0$\pm$2.0 &   3.0$\pm$1.2 &  $-$2.6$\pm$3.2 &  abs  \nl
  493 &   11.71 &   98.91 & 21.70 &     1.01 &  0.33082 &  0.00020 &   3.1$\pm$3.3 &   3.4$\pm$3.5 &   3.5$\pm$2.1 &  $-$1.2$\pm$3.2 &  abs  \nl
  498 &   67.82 &  106.81 & 21.28 &     0.98 &  0.33029 &  0.00020 &   4.3$\pm$6.0 &  $-$2.1$\pm$5.2 &   3.2$\pm$1.5 &  2.7$\pm$10.0 &  abs  \nl
  507 & $-$181.26 &  112.42 & 20.45 &   0.81 &  0.32387 &  0.00020 &   3.7$\pm$1.9 &   5.8$\pm$2.6 &   4.1$\pm$2.1 &  $-$1.7$\pm$2.5 & e+a \nl
  510 &   55.50 &  117.34 & 20.50 &     0.97 &  0.33322 &  0.00020 &   2.5$\pm$1.1 &   2.4$\pm$1.4 &   3.2$\pm$1.1 &  $-$1.4$\pm$2.1 &  abs  \nl
  514 &  $-$58.30 &  119.21 & 21.06 &   0.90 &  0.33070 &  0.00020 &   2.6$\pm$2.7 &   1.1$\pm$1.1 &   2.8$\pm$1.4 &   4.0$\pm$3.1 &  abs  \nl
  519 &  $-$32.42 &  124.38 & 20.84 &   0.66 &  0.32390 &  0.00020 &   0.2$\pm$1.5 &   0.3$\pm$1.7 & $-$12.3$\pm$1.7 & $-$41.1$\pm$3.6 & emi \nl
  522 & $-$125.36 &  124.35 & 21.44 &   0.90 &  0.33011 &  0.00020 &   3.5$\pm$4.2 &   0.0$\pm$5.8 &   4.8$\pm$4.4 &  $-$0.1$\pm$5.7 &  abs  \nl
  523 &   37.47 &  126.64 & 20.21 &     1.07 &  0.32323 &  0.00020 &   1.0$\pm$3.3 &  $-$1.5$\pm$4.5 &   3.3$\pm$1.8 &   1.0$\pm$2.8 &  abs  \nl
  525 & $-$137.72 &  125.54 & 20.48 &   0.96 &  0.33137 &  0.00020 &   1.1$\pm$2.5 &   0.3$\pm$1.9 &   3.3$\pm$1.1 &  $-$0.1$\pm$3.0 &  abs  \nl
  528 &  $-$17.83 &  132.87 & 20.78 &   0.55 &  0.32373 &  0.00020 &   4.9$\pm$1.6 &   2.5$\pm$1.8 &  $-$6.8$\pm$1.9 & $-$46.0$\pm$3.6 & emi+H$\delta$ \nl
  531 &  $-$68.80 &  136.31 & 19.21 &   1.02 &  0.33075 &  0.00020 &   0.5$\pm$1.0 &  $-$0.6$\pm$1.8 &   2.4$\pm$0.8 &   1.2$\pm$1.5 &  abs  \nl
  534 & $-$124.68 &  140.37 & 20.83 &   0.97 &  0.32356 &  0.00020 &  $-$1.3$\pm$3.8 &   2.2$\pm$2.8 &   2.1$\pm$1.9 &  $-$1.0$\pm$3.5 &  abs  \nl
  536 &   17.01 &  144.87 & 19.34 &     1.02 &  0.32789 &  0.00020 &  $-$0.2$\pm$1.0 &   0.6$\pm$1.3 &   2.1$\pm$0.9 &   1.1$\pm$1.6 &  abs  \nl
  537 &   31.69 &  145.16 & 20.66 &     0.98 &  0.32602 &  0.00020 &  $-$0.5$\pm$2.0 &   1.7$\pm$2.3 &   1.2$\pm$1.0 &  $-$0.2$\pm$1.8 &  abs  \nl
  539 &  108.68 &  146.75 & 21.20 &     1.03 &  0.32589 &  0.00020 &   1.7$\pm$4.1 &   3.1$\pm$3.9 &   2.2$\pm$1.7 &   0.5$\pm$6.1 &  abs  \nl
  540 &   68.26 &  147.10 & 20.87 &     1.00 &  0.32781 &  0.00020 &   2.3$\pm$2.7 &  $-$1.0$\pm$2.4 &   1.9$\pm$1.5 &  $-$3.3$\pm$3.9 &  abs  \nl
  542 &   $-$4.28 &  150.03 & 20.23 &   1.00 &  0.33162 &  0.00020 &   0.4$\pm$1.3 &   0.2$\pm$2.1 &   3.1$\pm$1.1 &   1.2$\pm$2.2 &  abs  \nl
  544 & $-$195.75 &  151.37 & 20.42 &   1.06 &  0.33280 &  0.00020 &  $-$0.3$\pm$2.0 &   0.0$\pm$2.2 &   2.6$\pm$1.3 &  $-$1.4$\pm$2.3 &  abs  \nl
  546 &  100.43 &  157.41 & 20.96 &     0.93 &  0.33118 &  0.00020 &  $-$0.4$\pm$3.1 &   2.7$\pm$2.3 &   2.8$\pm$2.5 &   2.6$\pm$2.7 &  abs  \nl
  549 &   47.65 &  161.11 & 20.55 &     0.91 &  0.33281 &  0.00020 &   2.6$\pm$2.2 &   2.3$\pm$2.3 &   2.7$\pm$2.0 &  $-$1.5$\pm$1.9 &  abs  \nl
  553 & $-$150.30 &  160.49 & 21.55 &   0.84 &  0.31894 &  0.00024 &   5.6$\pm$3.6 &   3.3$\pm$4.3 &   2.1$\pm$2.3 &  $-$2.4$\pm$6.0 &  abs  \nl
  554 & $-$252.31 &  160.29 & 19.95 &   1.00 &  0.32971 &  0.00020 &  $-$0.3$\pm$1.4 &   0.4$\pm$1.5 &   2.7$\pm$0.8 &  $-$0.3$\pm$1.6 &  abs  \nl
  555 &   82.82 &  164.56 & 21.48 &     1.00 &  0.32956 &  0.00020 &   3.1$\pm$2.0 &   0.5$\pm$3.1 &   3.4$\pm$1.8 &   1.6$\pm$5.3 &  abs  \nl
  560 &   21.67 &  170.53 & 20.41 &     0.99 &  0.33136 &  0.00020 &   1.6$\pm$1.1 &   0.5$\pm$1.9 &   2.4$\pm$0.9 &  $-$1.2$\pm$2.1 &  abs  \nl
  562 &   57.01 &  175.13 & 20.91 &     0.73 &  0.33643 &  0.00020 &   6.2$\pm$4.5 &   1.1$\pm$5.1 &   7.0$\pm$4.6 &   0.6$\pm$5.3 & e+a \nl
  565 &  140.21 &  177.65 & 21.78 &     0.80 &  0.32449 &  0.00030 &   5.6$\pm$4.2 &   4.4$\pm$4.5 &   4.7$\pm$3.5 &   2.5$\pm$5.2 & e+a \nl 
  572 & $-$198.72 &  181.74 & 20.25 &   0.94 &  0.33287 &  0.00020 &   1.4$\pm$0.8 &   0.3$\pm$2.3 &   1.9$\pm$1.0 &  $-$3.1$\pm$3.0 &  abs  \nl
  584 &  $-$54.21 &  201.30 & 21.36 &   0.97 &  0.32949 &  0.00033 &  $-$1.7$\pm$4.5 &   1.8$\pm$3.8 &   1.8$\pm$1.9 &  $-$2.2$\pm$9.9 &  abs  \nl
  587 &   43.66 &  206.78 & 21.61 &     0.82 &  0.33483 &  0.00035 &   4.1$\pm$3.2 &   2.4$\pm$2.0 &   2.0$\pm$2.4 &  $-$5.8$\pm$7.0 & abs \nl
  591 & $-$204.21 &  209.09 & 21.81 &   0.63 &  0.33451 &  0.00020 &   4.8$\pm$4.7 &   8.1$\pm$4.3 &  $-$1.7$\pm$5.0 & $-$18.1$\pm$9.1 & emi+H$\delta$ \nl
  594 & $-$169.55 &  213.66 & 21.85 &   0.70 &  0.33453 &  0.00032 &  10.3$\pm$2.0 &   4.5$\pm$1.1 &   3.6$\pm$3.0 &  $-$9.4$\pm$4.5 & emi+H$\delta$ \nl
  595 &  261.33 &  219.25 & 19.52 & \nodata &  0.33694 &  0.00020 &   4.1$\pm$0.7 &   1.6$\pm$0.8 &  $-$5.7$\pm$1.0 & $-$22.5$\pm$1.3 & emi+H$\delta$ \nl
  599 &  237.32 &  224.40 & 20.65 & \nodata &  0.33514 &  0.00020 &  $-$0.2$\pm$2.3 &   1.7$\pm$2.8 &   2.5$\pm$1.5 &   0.4$\pm$2.2 &  abs  \nl
  604 & $-$223.01 &  221.73 & 20.40 &   0.93 &  0.32494 &  0.00020 &   0.8$\pm$3.2 &   0.3$\pm$2.1 &   2.5$\pm$1.7 &  $-$1.4$\pm$4.1 &  abs  \nl
  626 &  $-$56.50 &  255.17 & 20.38 &   1.01 &  0.33207 &  0.00020 &   1.1$\pm$3.8 &  $-$0.3$\pm$3.7 &   2.8$\pm$2.1 &   2.0$\pm$4.5 &  abs  \nl
  636 &   31.61 &  270.99 & 20.83 & \nodata &  0.32433 &  0.00020 &   4.4$\pm$3.0 &   2.7$\pm$5.2 &   3.6$\pm$5.2 &  $-$8.4$\pm$3.5 & emi+H$\delta$ \nl
  647 &  140.31 &  280.70 & 20.24 &     0.94 &  0.33096 &  0.00020 &   2.4$\pm$2.1 &   1.6$\pm$1.7 &   3.0$\pm$1.1 &  $-$2.2$\pm$3.1 &  abs  \nl
  649 & $-$269.81 &  278.28 & 19.99 & \nodata &  0.32803 &  0.00020 &   0.9$\pm$2.5 &   0.1$\pm$2.8 &   3.3$\pm$1.4 &   2.6$\pm$3.6 &  abs  \nl
  652 &  205.97 &  286.74 & 21.39 & \nodata &  0.33304 &  0.00082 &   3.1$\pm$5.0 &   0.6$\pm$4.4 &  $-$0.2$\pm$2.6 & $-$24.5$\pm$6.5 & emi \nl
 1155 &  173.49 & $-$253.08 & 22.90 & \nodata &  0.33406 &  0.00020 & \nodata & \nodata & \nodata &$-$47.4$\pm$23.7 & emi \nl
 1157 &   35.80 & $-$254.31 & 21.83 & \nodata &  0.33328 &  0.00020 &  $-$2.1$\pm$2.8 &  $-$1.6$\pm$5.3 &   2.5$\pm$2.6 &  $-$3.1$\pm$7.0 &  abs  \nl
 1188 &  $-$88.63 & $-$238.14 & 22.08 & 0.75 &  0.32222 &  0.00029 &   3.9$\pm$6.4 &  $-$1.5$\pm$6.2 &  $-$3.5$\pm$4.2 &  $-$6.7$\pm$5.8 & emi \nl
 1195 &  108.86 & $-$228.36 & 22.90 &   0.33 &  0.32956 &  0.00021 &  11.5$\pm$11.9 &   3.2$\pm$15.5 &  $-$3.4$\pm$ 6.0 &$-$34.8$\pm$11.4 & emi \nl
 1328 & $-$192.34 & $-$159.57 & 21.86 & 1.00 &  0.33102 &  0.00020 &   2.3$\pm$3.4 &  $-$0.1$\pm$3.4 &   2.0$\pm$2.8 &   1.9$\pm$5.7 &  abs  \nl
 1341 & $-$242.25 & $-$150.00 & 22.40 & 0.51 &  0.32058 &  0.00047 &   5.1$\pm$5.3 &   2.9$\pm$5.8 &   0.3$\pm$4.2 & $-$30.7$\pm$7.5 & emi \nl
 1352 & $-$139.82 & $-$144.49 & 21.91 & 0.80 &  0.32833 &  0.00022 &   4.1$\pm$4.3 &   0.4$\pm$4.9 &   3.1$\pm$3.2 &   7.9$\pm$9.3 &  abs  \nl
 1414 &  116.29 & $-$115.11 & 21.98 & 0.94 &  0.33304 &  0.00020 &   3.9$\pm$5.1 &   4.8$\pm$7.4 &  $-$0.6$\pm$3.9 &  $-$1.6$\pm$6.1 &  abs  \nl
 1455 &   $-$3.46 &  $-$89.39 & 22.13 & 0.90 &  0.33028 &  0.00020 &  $-$2.2$\pm$5.4 &   2.2$\pm$5.3 &   2.8$\pm$3.1 &  $-$0.3$\pm$7.2 &  abs  \nl
 1461 &   97.04 &  $-$83.36 & 22.29 & 0.71 &  0.32747 &  0.00032 &  $-$3.4$\pm$9.7 &   4.0$\pm$7.3 &   6.2$\pm$5.8 &  7.7$\pm$14.1 &  abs  \nl
 1475 &  $-$19.94 &  $-$76.88 & 20.57 & \nodata &  0.33191 &  0.00020 &   2.0$\pm$1.6 &   2.6$\pm$2.5 &   2.2$\pm$1.1 &   2.1$\pm$1.6 &  abs  \nl
 1487 &  139.12 &  $-$58.63 & 23.08 & 0.41 &  0.32081 &  0.00020 &  10.6$\pm$ 15.5 &   3.7$\pm$17.4 & $-$19.9$\pm$30.7 & $-$43.9$\pm$8.8 & emi \nl
 1524 &   82.16 &  $-$39.65 & 21.86 & 0.98 &  0.33574 &  0.00020 &  $-$0.5$\pm$3.5 &   2.2$\pm$3.7 &   1.3$\pm$4.1 &  $-$0.6$\pm$4.4 &  abs  \nl
 1563 &  $-$62.95 &  $-$23.67 & 21.47 & \nodata &  0.33136 &  0.00020 &  $-$0.2$\pm$ 3.8 &  $-$3.7$\pm$5.4 &   3.5$\pm$2.3 &   3.1$\pm$7.7 &  abs  \nl
 1594 &  $-$21.42 &  $-$14.73 & 22.04 & \nodata &  0.32330 &  0.00024 &  $-$4.8$\pm$15.6 & $-$18.6$\pm$19.2 &   6.2$\pm$13.6 &  3.0$\pm$10.9 &  abs  \nl
 1604 &   95.36 &   $-$5.09 & 22.54 & 0.91 &  0.32662 &  0.00027 &   4.5$\pm$ 14.2 &  $-$7.2$\pm$12.8 &   1.4$\pm$ 7.2 & $-$2.9$\pm$18.4 &  abs  \nl
 1616 &   30.67 &   $-$1.91 & 22.02 & 1.01 &  0.33368 &  0.00020 &   2.9$\pm$2.9 &   3.7$\pm$4.3 &   2.4$\pm$2.6 &   3.8$\pm$4.9 &  abs  \nl
 1630 &   92.41 &   12.82 & 22.15 & 0.62 &  0.33221 &  0.00020 &   8.2$\pm$4.6 &  $-$3.5$\pm$6.0 &  $-$5.2$\pm$2.8 & $-$29.3$\pm$7.5 & emi+H$\delta$ \nl
 1649 &   71.56 &   20.50 & 21.62 & 0.97 &  0.32544 &  0.00020 &   2.4$\pm$3.7 &   2.0$\pm$3.1 &   1.3$\pm$4.0 &   3.2$\pm$3.2 &  abs  \nl
 1731 &  $-$75.92 &   67.56 & 21.84 & 0.82 &  0.32830 &  0.00020 &  $-$0.8$\pm$3.6 &   3.7$\pm$2.7 &   2.5$\pm$2.4 &   3.5$\pm$3.6 &  abs  \nl
 1775 &  116.86 &   97.78 & 21.65 & 0.88 &  0.33430 &  0.00027 &  10.5$\pm$7.6 &   2.5$\pm$9.4 &  $-$1.9$\pm$6.2 &  $-$0.1$\pm$9.9 &  abs  \nl
 1806 &   93.60 &  118.92 & 21.75 & 1.08 &  0.33190 &  0.00020 &   3.4$\pm$3.0 &  $-$9.0$\pm$4.3 &   1.3$\pm$3.6 &   2.9$\pm$7.7 &  abs  \nl
 1810 &  275.48 &  122.77 & 22.21 & \nodata &  0.32758 &  0.00023 &   6.5$\pm$ 10.6 &  $-$8.5$\pm$8.2 &  $-$8.4$\pm$5.2 &$-$56.6$\pm$12.4 & emi \nl
 1816 &   77.81 &  123.90 & 22.21 & 0.82 &  0.33413 &  0.00020 &  $-$1.0$\pm$4.2 &   2.6$\pm$4.3 &   0.0$\pm$4.3 & $-$3.8$\pm$14.7 &  abs  \nl
 1829 &  136.67 &  131.11 & 23.03 & 0.76 &  0.33386 &  0.00020 &   2.4$\pm$ 13.2 &  $-$0.5$\pm$9.6 &  $-$4.1$\pm$10.8 &$-$27.6$\pm$25.9 & emi \nl
 1842 &  $-$14.42 &  133.77 & 22.09 & 0.86 &  0.32299 &  0.00024 &   4.1$\pm$6.2 &  $-$2.7$\pm$6.5 &   0.0$\pm$5.2 &  $-$4.2$\pm$9.3 &  abs  \nl
 1859 &  $-$76.77 &  142.06 & 21.99 & 0.96 &  0.33254 &  0.00020 &   2.6$\pm$4.1 &   3.2$\pm$2.5 &   2.6$\pm$4.1 &   3.8$\pm$5.4 &  abs  \nl
 1865 &  111.08 &  146.35 & 21.43 & \nodata &  0.32847 &  0.00020 &  $-$1.2$\pm$8.8 &  $-$2.8$\pm$9.0 &   4.8$\pm$6.0 & $-$5.0$\pm$10.7 & abs \nl
 1871 &   16.32 &  147.17 & 20.37 & \nodata &  0.32681 &  0.00020 &   0.1$\pm$1.7 &   0.8$\pm$2.1 &   2.6$\pm$1.0 &   0.3$\pm$1.5 &  abs  \nl
 1897 &  138.96 &  166.83 & 21.67 & 0.79 &  0.33097 &  0.00024 &   2.7$\pm$7.0 &   1.9$\pm$4.6 &   2.3$\pm$3.4 & $-$0.3$\pm$14.4 &  abs  \nl
 1978 &  140.46 &  229.01 & 22.31 & 1.04 &  0.33076 &  0.00024 &   2.1$\pm$9.9 &   3.3$\pm$10.8 &   2.1$\pm$7.4 &$-$11.1$\pm$29.8 & abs \nl
 2081 & $-$103.05 &  281.87 & 22.36 & 0.45 &  0.33467 &  0.00077 &  $-$8.8$\pm$18.7 &   1.9$\pm$13.8 & $-$14.0$\pm$20.7 &$-$65.4$\pm$22.3 & emi \nl

\enddata
\end{deluxetable}

\begin{deluxetable}{cccc}
\small
\tablewidth{0pc}
\tablenum{2}
\tablecaption{Line Strength Index Definitions}
\tablehead{
\colhead{Index} &
\colhead{Bandpass} &
\colhead{Blue sideband} &
\colhead{Red sideband}
}
\startdata
H$\delta$ & 4083.500$-$4122.250 & 4017.00$-$4057.00 & 4153.00$-$4193.00 \nl
H$\gamma$ & 4319.750$-$4363.500 & 4242.00$-$4282.00 & 4404.00$-$4444.00 \nl
H$\beta$  & 4847.875$-$4876.625 & 4799.00$-$4839.00 & 4886.00$-$4926.00 \nl
[OII]     & 3716.300$-$3738.300 & 3696.30$-$3716.30 & 3738.30$-$3758.30 \nl
\enddata
\end{deluxetable}

\begin{deluxetable}{lccc}
\small
\tablewidth{0pc}
\tablenum{3}
\tablecaption{Spectral Type Kinematics}
\tablehead{
\colhead{Type} &
\colhead{Num} &
\colhead{Mean} &
\colhead{Dispersion}
}
\startdata
All            & 232 & 98425$\pm$90\phn  & 1027$^{+51}_{-45}$\phm{$^2$}  \nl
Absorption     & 177 & 98464$\pm$97\phn  & \phn972$^{+56}_{-48}$\phm{$^2$} \nl
E+A            & \phn11 & 97891$\pm$567 & 1414$^{+463}_{-234}$ \nl
Emission       & \phn44 & 98402$\pm$228 & 1136$^{+146}_{-105}$ \nl
Emi ($<$20\AA) & \phn23 & 98419$\pm$303 & 1092$^{+211}_{-134}$ \nl
Emi ($>$20\AA) & \phn21 & 98383$\pm$352 & 1210$^{+248}_{-154}$ \nl
Emi+H$\delta$  & \phn16 & 98059$\pm$419 & 1260$^{+312}_{-179}$ \nl
\tablecomments{Kinematic parameters as a function of spectral type.
The emission-line galaxy spectra have been subdivided into those
showing moderate (5\thinspace\AA$\leq$ [OII]
3727\thinspace\AA$\leq$20\thinspace\AA) and strong ([OII]
3727\thinspace\AA$\geq$20\thinspace\AA) emission and those with
strong H$\delta$ absorption ($\geq$4\thinspace\AA) irrespective of
[OII] strength.}
\enddata  
\end{deluxetable}   

\end{document}